\documentclass[runningheads]{llncs}
\usepackage{etoolbox}
\AtBeginEnvironment{theorem}{\vspace{-2pt}}
\AtEndEnvironment{theorem}{\vspace{-2pt}}
\AtBeginEnvironment{lemma}{\vspace{-2pt}}
\AtEndEnvironment{lemma}{\vspace{-2pt}}
\AtBeginEnvironment{proposition}{\vspace{-2pt}}
\AtEndEnvironment{proposition}{\vspace{-3pt}}
\AtBeginEnvironment{definition}{\vspace{-3.3pt}}
\AtEndEnvironment{definition}{\vspace{-3.3pt}}
\AtBeginEnvironment{remark}{\vspace{-3pt}}
\AtEndEnvironment{remark}{\vspace{-3pt}}
\usepackage{enumitem}
\setlist{topsep=0pt, itemsep=0pt, parsep=0pt}
\makeatletter
\renewcommand\section{\@startsection{section}{1}{\z@}%
                       {-3.3ex \@plus -0.5ex \@minus -.2ex}%
                       {0.8ex \@plus .2ex}%
                       {\normalfont\large\bfseries}}

\renewcommand\subsection{\@startsection{subsection}{2}{\z@}%
                          {-3.3ex \@plus -0.3ex \@minus -.2ex}%
                          {0.5ex \@plus .2ex}%
                          {\normalfont\normalsize\bfseries}}
\makeatother

\usepackage[T1]{fontenc}

\usepackage{float}
\usepackage{thm-restate}
\usepackage{algorithm}
\usepackage{algpseudocode}
\usepackage[dvipsnames]{xcolor}
\usepackage{verbatim}
\usepackage{etoolbox}
\usepackage{thmtools}
\newtoggle{showproofs}
\settoggle{showproofs}{true}   

\iftoggle{showproofs}{
}{
  \renewenvironment{proof}{\comment}{\endcomment}
}

\algrenewcommand{\algorithmicrequire}{\textbf{Input:}}
\algrenewcommand{\algorithmicensure}{\textbf{Output:}}

\newcommand{\Dist}{\mathrm{Dist}}
\newcommand{\Post}{\mathrm{Post}}
\newcommand{\Pred}{\mathrm{Pred}}
\newcommand{\Nat}{\mathbb{N}}
\newcommand{\Prob}{\mathbb{P}}
\newcommand{\Exp}{\mathbb{E}}

\usepackage{newtxtext}       %
\usepackage[varvw]{newtxmath}       

\usepackage{graphicx}

\usepackage{xcolor}
\usepackage[
    pagebackref=false,
    breaklinks=true,
    colorlinks=true,
    linkcolor=blue,
    citecolor=blue,
    urlcolor=blue
]{hyperref}

\begin{document}
\title{Quantitative Reachability Properties Beyond Decisiveness}

\author{Raphaël Faure\inst{1} \and
Alain Finkel\inst{2} \and
Lina Ye\inst{3}}

\authorrunning{R. Faure et al.} 

\institute{Universit\'e Paris-Saclay,  CNRS, 
ENS Paris-Saclay, Laboratoire Méthodes Formelles, 
91190, Gif-sur-Yvette, France\\ \email{raphaelfaure@lmf.cnrs.fr}\\
\and
Universit\'e Paris-Saclay, CNRS, 
ENS Paris-Saclay, Laboratoire Méthodes Formelles,  
91190, Gif-sur-Yvette, France\\
\email{alain.finkel@ens-paris-saclay.fr}\\
 \and
Universit\'e Paris-Saclay, CNRS, 
ENS Paris-Saclay, CentraleSupélec, Laboratoire Méthodes Formelles,  
91190, Gif-sur-Yvette, France\\
\email{lina.ye@centralesupelec.fr}}

\maketitle              
\begin{abstract}
    Decisiveness is central to reachability algorithms for infinite-state
Markov chains, but is a qualitative notion: it guarantees that
path-enumeration algorithms terminate, without an \emph{a priori} bound on
the horizon they require. In this work, we study decisiveness from a quantitative
point of view at the level of Markov chains, and apply our results to the
approximate quantitative coverability problem for effective probabilistic
well-structured transition systems (pWSTS). We introduce truncation sets, which represent the horizons sufficient to approximate the reachability probability within a precision $\varepsilon$. These sets reformulate decisiveness quantitatively, and their elements can be computed by exploring the chain when it is decisive. To obtain such horizons \emph{a priori}, before any
exploration, we introduce geometrically reachable sets, for which the probability of not reaching them within $n$ steps decreases exponentially in $n$, up to a prefactor depending on the initial state.
Horizons valid for all initial states exist if and only if this prefactor
can be chosen uniformly, that is, for boundedly coarse chains. For such
effective pWSTS, including probabilistic vector addition systems and
Galton--Watson processes, we obtain \emph{a priori} global horizons
growing at most logarithmically in $1/\varepsilon$. In the non-uniform
case, we give a drift condition ensuring the existence of a finite
geometrically reachable set. Under this condition, we derive
\emph{a priori} horizons growing at most logarithmically in
$1/\varepsilon$ and in the weight $w(x)$ measuring the size of the
initial state $x$. In both cases, we compute, once for each target, a
horizon function whose evaluation on any initial state and precision
requires only a constant number of elementary operations
$(+, -, \times, / \text{ and } \log)$. For probabilistic lossy channel systems, this
yields, to our knowledge, the first explicit \emph{a priori} horizons
from arbitrary initial states, and it extends to $\nu$-pLCS, a new
infinitely branching variant with probabilistic emissions.

\end{abstract}

\section{Introduction}
\label{sec:intro}


\paragraph*{General context.}
Quantitative verification of infinite-state Markov chains often reduces to
approximating the probability of reaching a target set. A standard approach
explores the paths of the chain up to some horizon and bounds the probability
mass that remains unexplored. Decisiveness~\cite{AbdullaBenHendaMayr2007} is
central to this approach: a Markov chain is decisive with respect to a
target if, with probability one, it eventually reaches either the target or
a state from which the target is unreachable. In particular, decisiveness guarantees that
the unexplored mass tends to zero, hence that path exploration terminates
for any precision. It is, however, a qualitative notion: it gives no
information on the horizon that the exploration requires, and this horizon is only
known once the exploration is over.

\vspace{-6pt}
\paragraph*{Decisiveness and its uses.}
Path-enumeration algorithms approximating reachability probabilities were
developed for probabilistic lossy channel systems
(pLCS)~\cite{IyerNarasimha1997}, and formulated for general decisive Markov
chains by Abdulla, Ben Henda and Mayr~\cite{AbdullaBenHendaMayr2007}, who
showed that Markov chains with a finite attractor, such as those induced by
pLCS, and globally coarse Markov chains, such as those induced by
probabilistic vector addition systems with states (pVASS), are decisive. Bertrand, Bouyer, Brihaye and
Carlier~\cite{bertrand_et_al:LIPIcs.ICALP.2016.101} extended decisiveness
to stochastic processes over general state spaces. Further works go beyond
decisiveness~\cite{FinkelHaddadYeDynamic2023,BarbotBouyerHaddad2024}. In
particular, a complementary property called divergence handles transient
systems by identifying regions that can be safely
truncated~\cite{FinkelHaddadYeDynamic2023}. Deciding decisiveness for
high-level formalisms has also been studied~\cite{FougeaHYJF25}. Finally,
probabilistic well-structured transition systems (pWSTS)~\cite{fffy2026}
provide a unified framework for Markov chains whose underlying transition
system is well structured, including probabilistic vector addition systems (pVAS), pLCS and
multi-type Galton--Watson processes. They extend path-enumeration
algorithms to effective pWSTS without any finite-branching assumption. In
all these works, decisiveness ensures termination, but none of them gives
a sufficient horizon to explore before the exploration starts.

\vspace{-6pt}
\paragraph*{Eagerness.}
The closest quantitative notion is eagerness, introduced by Abdulla,
Ben Henda, Mayr and Sandberg~\cite{AbdullaBenHendaMayrSandbergATVA2006,AbdullaBenHendaMayrSandbergQEST2006,BenHenda08}. A Markov chain is eager with respect to an initial state if the probability
that the target is \emph{eventually} reached, but only after $n$ or more
steps, decreases exponentially in $n$.
Eagerness ensures that a path-enumeration algorithm  approximates expected
rewards up to any precision. 
However, eagerness is not a quantitative form of decisiveness: the measured
events are intersected with the event that the target is eventually reached. A chain
is thus trivially eager with respect to a state from which the target is
unreachable, and may be eager without being decisive. 

\vspace{-6pt}
\paragraph*{Our approach.}
We aim to study decisiveness from a quantitative point of view. For a fixed target, we seek an \textit{a priori} function mapping any initial state and desired precision to a sufficient exploration horizon. We formally define these functions at the Markov chain level and compute them for effective pWSTS in the context of approximate quantitative coverability. 

\vspace{-6pt}
\paragraph*{Contributions.}
Our main contributions are fourfold:
\begin{itemize}
  \item \emph{Quantitative decisiveness.} We introduce \emph{truncation
  sets}, the horizons sufficient for an $\varepsilon$-approximation of the
  probability of reaching a target $A$, and show that they are all
  non-empty if and only if the chain is decisive with respect to A, in which case their elements can be computed by exploring the chain. We then
  introduce \emph{geometrically reachable sets} for which the probability of not reaching them within $n$ steps decreases exponentially in $n$. When $A$ and the states from which $A$ is unreachable form such a
  set, we obtain an \emph{a priori} sufficient horizon, hence a quantitative
  form of decisiveness. Unlike eagerness,
  this notion implies decisiveness.

  \item \emph{Uniform bounds.} We show that a single horizon valid for all
  initial states exists if and only if the chain is boundedly coarse, or
  equivalently when the geometric bound is uniform in the initial state.
  For boundedly coarse effective pWSTS, we compute with the backward
  coverability algorithm an \textit{a priori} horizon function in
  $O(\log(1/\varepsilon))$ of the precision $\varepsilon$, whose
  evaluation requires a constant (independent of $\varepsilon$) number of elementary operations.

  \item \emph{Initial-state-dependent bounds.} We show that, given a finite
  geometrically reachable set, the chain reaches any target, or a state
  from which it is unreachable, geometrically fast from every initial
  state. We give a drift condition on a weight function $w$ ensuring the
existence of such a set, which also applies to infinitely branching
chains. Under this condition, it yields an
  \textit{a priori} horizon function in $O(\log(1/\varepsilon) + \log w(x))$
  of the precision $\varepsilon$ and the initial state $x$, which is computable for effective pWSTS and whose evaluation requires a constant (independent of $\varepsilon$ and $x$) number of
  elementary operations once $w(x)$ is known.

   \item \emph{Applications.} We instantiate the uniform case on pVAS and
  multi-type Galton--Watson processes, and the non-uniform case on pLCS
  and on $\nu$-pLCS, a new model of lossy channel systems whose emissions
  append a random number of messages, and which may be infinitely
  branching. To the best of our knowledge, such explicit horizons from
  arbitrary initial states had not been stated before for any of these
  models. 
\end{itemize}

\paragraph*{Roadmap.}
Section~\ref{sec:preli} recalls WSTS, Markov chains and pWSTS.
Section~\ref{sec:geo-reach} introduces truncation sets, the procedures
computing their elements, and geometrically reachable sets.
Section~\ref{sec:coarseness} studies uniformly geometrically reachable
sets, their link with bounded coarseness, and their application to pVAS
and Galton--Watson processes. Section~\ref{sec:beyond-coarseness} treats
the non-uniform case, builds finite geometrically reachable sets from
drift conditions, and applies them to pLCS and $\nu$-pLCS.
Section~\ref{sec:conclusion} concludes, and a statement on the use of AI
opens the appendix.

\section{Preliminaries}
\label{sec:preli}

\subsection{Transition systems}

A \emph{transition system} is a pair $(S, \to)$, where $S$ is a set of states and ${\to} \subseteq S \times S$ is a transition relation. A \emph{run} of $(S, \to)$ is a (finite or infinite) sequence $s_0 \rightarrow s_1 \rightarrow \cdots$ such that $s_i\rightarrow s_{i+1}$ for every consecutive pair. 
In the literature, runs are also referred to as \emph{paths}, which explains the usual name \emph{path-enumeration algorithm}. We write $s \rightarrow^{m} t$ when there is a run of length $m \in \Nat$ from $s$ to $t$, with the convention $s \rightarrow^{0} s$, and $\rightarrow^{*}$ denotes the reflexive transitive closure of $\rightarrow$. From now on, the run $s_0 \rightarrow s_1 \rightarrow \cdots$ is written simply as $s_0s_1...$ . Given $X \subseteq S$, the predecessor operators are defined as follows:
\begin{itemize}
  \item for $m\in \Nat$, $\Pred^m(X) = \{\, s \in S \mid \exists\, t \in X,\ s \to^m t \,\}$; in particular, $\Pred^0(X) = X$, and we write $\Pred(X)$ for $\Pred^1(X)$;
  \item for $\ell\in \Nat$, $\Pred^{\leq \ell}(X) = \bigcup_{0 \leq m \leq \ell} \Pred^m(X)$, and $\Pred^*(X) = \bigcup_{m \geq 0} \Pred^m(X)$.
\end{itemize}
The successor operators $\Post^m(X)$, $\Post^{\le \ell}(X)$ and $\Post^{*}(X)$ are defined in the same way. For a single element $x \in S$, we abbreviate $\Pred(\{x\})$ as $\Pred(x)$ and $\Post(\{x\})$ as $\Post(x)$, and similarly for the iterated operators.

We now add an order structure on states. Let $S$ be a set. A \emph{quasi-order} (qo) on $S$ is a binary relation $\leq$ on $S$ that is both reflexive and transitive. Given $(S, \le)$, the upward-closure of a set $X\subseteq S$ is $\uparrow X=\{y\in S \mid \exists x\in X,\ x\le y\}$, and for $x\in S$ we set $\uparrow x=\uparrow\{x\}$. A set $U\subseteq S$ is \emph{upward-closed} when $U=\uparrow U$. We say that $\le$ is a \emph{well-quasi-order} (wqo) if, for every infinite sequence $(x_i)_{i \in \Nat}$ of elements of $S$, there exist indices $i < j$ such that $x_i \le x_j$. A classical and fundamental property of wqo states that every upward-closed set $U\subseteq S$ has a finite basis, that is, a finite set $B\subseteq U$ such that $U=\uparrow B$.

Given $(S, \le)$, a transition relation ${\to} \subseteq S \times S$ is called \emph{strongly monotone} if for all $x, y, x' \in S$:
$
  \bigl(x \le x' \ \wedge\ x \to y\bigr) \ \Longrightarrow\ \bigl(\exists y' \in S :\ x' \to y' \ \wedge\ y \le y'\bigr).
$
Strong monotonicity ensures that $\Pred(X)$ is upward-closed whenever $X$ is. Consequently, $\Pred^m(X)$ for every $m\in\mathbb{N}$, as well as $\Pred^*(X)$, are upward-closed too.

\begin{definition}[Well-Structured Transition System \cite{Finkel87,FinkelSchnoebelen,AbdullaCeransJonssonTsay2000}]
\label{def:wsts}
A \emph{Well-Structured Transition System with strong monotonicity} (WSTS) is a tuple $\mathcal{S}=(S,\le,\to)$, where: $(S, \to)$ is a transition system, $\le$ is a well-quasi-order on $S$, $\to$ is strongly monotone.
\end{definition}

The terminology Well-Structured Transition System encompasses several closely related but not fully equivalent definitions in the literature, which differ notably in their monotonicity condition (see \cite{FinkelSchnoebelen} for an overview). Throughout this paper, WSTS refers to the strongly monotone version defined above. We use this notion as the underlying structure of probabilistic WSTS (pWSTS)~\cite{fffy2026}, which are recalled below.

\begin{lemma}\cite{AbdullaCeransJonssonTsay2000}
\label{lem:finite-depth}
Let $\mathcal S=(S,\le,\to)$ be a WSTS and $U\subseteq S$ an upward-closed set.
There exists $\ell\in\Nat$ such that $\Pred^*(U)=\Pred^{\le \ell}(U)$.
\end{lemma}

We also recall the notion of Pred-Basis effectiveness.

\begin{definition}\cite{AbdullaCeransJonssonTsay2000}
\label{def:effective-wsts}
A WSTS $(S,\le,\to)$ is \emph{Pred-Basis effective} if $\le$ is decidable and, given any finite set $B\subseteq S$, one can effectively compute a finite basis of $\mathrm{Pred}(\uparrow B)$.
\end{definition}

Given a Pred-Basis effective WSTS and a finite set $B$, Lemma~\ref{lem:finite-depth} guarantees that a finite basis of $\Pred^*(\uparrow B)$ can be computed with the backward coverability algorithm~\cite{Abdulla_1996,FinkelSchnoebelen}.

\subsection{Markov chains and probabilistic WSTS}

In this paper, every countable set $S$ is equipped with the $\sigma$-algebra $2^S$. We write $\Dist(S)$ for the set of discrete probability distributions over $S$, and $\mathrm{supp}(\mu) = \{ x \in S \mid \mu(x) > 0 \}$ for the support of $\mu \in \Dist(S)$, where $\mu(x)$ abbreviates $\mu(\{x\})$. A \emph{Markov kernel} on $S$ is a mapping $P \colon S \to \Dist(S)$.

Let $P$ be a Markov kernel on a countable set $S$, $\Omega = S^{\mathbb{N}}$ the space of infinite runs, $X_k$ the $k$-th coordinate projection, $\mathcal{F}_n = \sigma(X_0, \dots, X_n)$ the natural filtration and $\mathcal{F} = \sigma(X_k : k \geq 0)$. For every $x \in S$, the Ionescu--Tulcea theorem~\cite{IonescuTulcea1949,Kallenberg2021} yields a unique probability measure $\mathbb{P}_x$ on $(\Omega, \mathcal{F})$ such that $X_0 = x$ and $\mathbb{P}_x(X_{n+1} = y \mid \mathcal{F}_n) = P(X_n)(y)$, $\mathbb{P}_x$-almost surely, for all $n \geq 0$ and $y \in S$. The process $(X_n)_{n \geq 0}$ is the Markov chain induced by $P$. The \emph{shift operator} $\theta_n \colon \Omega \to \Omega$, defined by $\theta_n((x_i)_{i \geq 0}) = (x_{n+i})_{i \geq 0}$, satisfies $X_k \circ \theta_n = X_{k+n}$.

A random variable $\tau \colon \Omega \to \mathbb{N} \cup \{\infty\}$ is a \emph{stopping time} if $\{\tau \leq n\} \in \mathcal{F}_n$ for every $n \geq 0$. For $A \subseteq S$, the \emph{hitting time} $\tau_A = \inf\{n \geq 0 : X_n \in A\}$ and the \emph{return time} $\tau_A^+ = \inf\{n \geq 1 : X_n \in A\}$ (with $\inf \emptyset = \infty$) are stopping times, related by $\tau_A^+ = 1 + \tau_A \circ \theta_1$. Moreover, for a stopping time $\tau$, the \emph{stopped $\sigma$-algebra} is $\mathcal{F}_\tau = \{ A \in \mathcal{F} \mid \forall n \geq 0,\ A \cap \{\tau = n\} \in \mathcal{F}_n \}$.\label{def:hitting-return}
\begin{proposition}[Markov properties]
\label{prop:markov}
Let $x \in S$ and $\Phi \colon \Omega \to \mathbb{R}$ be bounded or
non-negative measurable. Then, $\mathbb{P}_x$-almost surely:
\begin{enumerate}
  \item (\emph{weak Markov property}) for every $n \geq 0$,
  $\mathbb{E}_x\big[\Phi \circ \theta_n \mid \mathcal{F}_n\big]
  = \mathbb{E}_{X_n}[\Phi]$;
  \item (\emph{strong Markov property}) for every stopping time $\tau$,
  $\mathbb{E}_x\big[\mathbf{1}_{\{\tau < \infty\}}\,(\Phi \circ \theta_\tau)
  \mid \mathcal{F}_\tau\big]
  = \mathbf{1}_{\{\tau < \infty\}}\, \mathbb{E}_{X_\tau}[\Phi]$.
\end{enumerate}
\end{proposition}
Decisive Markov chains were introduced by Abdulla et al.~\cite{AbdullaBenHendaMayr2007} and further studied in~\cite{FinkelHaddadYeDynamic2023,BarbotBouyerHaddad2024,bertrand_et_al:LIPIcs.ICALP.2016.101}. From $P$, one can define a transition relation on $S$ by $x \to y$ iff
$P(x)(y) > 0$, yielding the transition system $(S, \to)$. For $A \subseteq S$, let $\widetilde A = S \setminus \Pred^*(A)$ be the set of states from which $A$ is unreachable. The Markov chain induced by $(S,P)$ is \emph{decisive with respect to $A$} if, from every initial state, almost every run eventually reaches $A$ or $\widetilde A$. Equivalently, the Markov chain is decisive with respect to $A$ if and only if $\mathbb{P}_x(\tau_{A \cup \widetilde A} < \infty) = 1$ for every $x \in S$. Finally, recall that a state $x \in S$ is a \emph{deadlock} if $\Post(x) = \emptyset$.

\begin{definition}[Probabilistic WSTS \cite{fffy2026}]
\label{def:pWSTS}
A \emph{probabilistic well-structured transition system (pWSTS)} is a pair $(\mathcal{S}, P)$, where $\mathcal{S} = (S, \leq, \to)$ is a WSTS and $P$ is a Markov kernel on $S$ that is compatible with the transition relation, in the following sense:
\begin{itemize}
    \item for every non-deadlock state $x \in S$ and every $y \in S$, $P(x)(y) > 0 \Longleftrightarrow x \to y$;
    \item for every deadlock state $x \in S$, $P(x)(x) = 1$.
\end{itemize}
\end{definition}

In Appendix~\ref{annexe-pWSTS-def-models}, we recall the definitions of pVAS (Definition~\ref{def:VAS}), pLCS under a \emph{loss-action} semantics and without deadlocks (Definition~\ref{def:pLCS}), and multi-type Galton--Watson processes (Definition~\ref{def:Galton--Watson}). These are the definitions used in \cite{fffy2026}, under which each of these models naturally induces a pWSTS. In what follows, we say that a pWSTS is decisive, boundedly coarse, and so on, when the Markov chain it induces is. 

\subsection{Effectiveness}

Let us recall that the kernel $P$ is \emph{computable}~\cite{fffy2026} if there exists an algorithm that, given $x, y \in S$ and a rational precision $\varepsilon > 0$, outputs rationals $a, b$ such that $a \leq P(x)(y) \leq b$ and $b - a \leq \varepsilon$. A mapping $\chi : S \to 2^S$, such as $\Post$, is \emph{effectively enumerable} if there exists a procedure $A$ that, on input $x \in S$ and an index $m \in \Nat_+$, either outputs an element of $\chi(x)$ or does not terminate, and such that: (i) every element of $\chi(x)$ is output for at least one index $m$; and (ii) $A$ may fail to terminate on an index $m$ only if the elements output for the indices $1, \dots, m-1$ already exhaust $\chi(x)$.

\begin{definition}[Effective pWSTS \cite{fffy2026}]
A pWSTS $(\mathcal{S}, P)$ is \emph{effective} if:
\begin{itemize}
    \item $\to$ is decidable;
    \item $=$ is decidable;
    \item $\mathcal{S}$ is a Pred-Basis effective WSTS;
    \item $\Post$ is effectively enumerable;
    \item $P$ is computable.
\end{itemize}
\end{definition}

Throughout this paper, we distinguish \emph{structural} results, stated for arbitrary Markov chains and independent of any effectiveness assumption, from their \emph{effective} counterparts, namely the computability of the parameters and functions these results provide, which requires an effective class. The procedures of~\cite{fffy2026} on which ours rely were designed for
effective pWSTS, and this is the class we consider in this work. We
denote an effective pWSTS by $(\mathcal{S}, P)_{\mathrm{eff}}$ in the sequel. We observe, however, that two of them, $\mathcal{A}_2$ and $\mathcal{SA}_3$ below, extend without effort to any Markov chain equipped with a target set $F$ such that $F$ has a finite representation, membership in $F$ and in $\Pred^*(F)$ is decidable, $P$ is computable, and the successors of each state are given by a function $\mathrm{succ}$ that effectively enumerates them without repetition. It suffices to replace membership in $\uparrow B$ and in $\Pred^*(\uparrow B)$ by membership in $F$ and in $\Pred^*(F)$. Their termination and correctness proofs remain unchanged, since they use neither the upward-closedness of the target nor the well-structured nature of the system \cite{fffy2026}. 

We call \emph{procedure} any computational scheme, without
implying, at this stage, termination. A procedure is a
\emph{semi-algorithm} if it terminates under additional conditions, and
an \emph{algorithm} if it terminates on every input. The three procedures of~\cite{fffy2026} used below are recalled in Appendix~\ref{annexe-proc}. The algorithm $\mathcal{A}_1$ computes, when $\uparrow B$ is reachable from $x$, a shortest run from $x$ to $\uparrow B$ together with its probability. The other two address the following problem.

\smallskip
\noindent\textbf{Approximate quantitative coverability problem (AQCP)~\cite{fffy2026}.}

\emph{Input:} $(\mathcal{S}, P)_{\mathrm{eff}}$, a finite $B \subseteq S$, $x \in S$, $\ell \in \Nat \cup \{\infty\}$ and $\varepsilon \in \mathbb{Q} \cap (0,1)$.

\emph{Output:} $p^\ast \in \mathbb{Q}$ such that $p^\ast \leq \mathbb{P}_x(\tau_{\uparrow B} < \ell + 1) \leq p^\ast + \varepsilon$ (with $\infty + 1 = \infty$).
\smallskip

The algorithm $\mathcal{A}_2$ solves the AQCP for every finite horizon $\ell$. The semi-algorithm $\mathcal{SA}_3$ solves it for $\ell = \infty$ when the chain is decisive with respect to $\uparrow B$, with no a priori bound on the horizon it explores.

\section{Geometric reachability}
\label{sec:geo-reach}

\subsection{Quantitative reachability}

We introduce \emph{truncation sets}, the horizons after which the
probability that the chain has reached neither the target nor a state
from which the target is unreachable is at most a prescribed precision.
This notion serves as the basis for the quantitative results of this
paper.

\begin{definition}[$\varepsilon$-truncation set]
\label{def:n-star}
Let $(S, P)$ be a Markov chain and $A \subseteq S$ a target subset.
For any $\varepsilon \in (0,1)$ and any state $x \in S$, the \emph{$\varepsilon$-truncation set} from $x$ for the target $A$ is defined as:
$
  n_A^*(\varepsilon, x) = \left\{ n \in \mathbb{N} :  
  \mathbb{P}_x(\tau_{A \cup \widetilde{A}} > n) \le \varepsilon \right\}. 
$
The global $\varepsilon$-truncation set over $S$ is defined as the intersection over all states:
$
  n_A^*(\varepsilon) = \bigcap_{x \in S} n_A^*(\varepsilon, x). 
$
\end{definition}

\begin{restatable}{proposition}{PropNStarProperties}
\label{prop:n-star-properties}
Let $(S, P)$ be a Markov chain, $A \subseteq S$, $(\varepsilon, x)\in (0,1)\times S$.
\begin{enumerate}
  \item The sets $n_A^*(\varepsilon, x)$ and $n_A^*(\varepsilon)$ are upward-closed in $\Nat$.
  \item The maps $\varepsilon \mapsto n_A^*(\varepsilon, x)$ and $\varepsilon \mapsto n_A^*(\varepsilon)$ are non-decreasing w.r.t.\ inclusion.
\end{enumerate}
\end{restatable}

\begin{proof}
\begin{enumerate}
    \item Since the events $\{\tau_{A\cup\widetilde A} > n\}$ are non-increasing in $n$, so is the map $n \mapsto \mathbb{P}_x(\tau_{A\cup\widetilde A} > n)$. Hence, if $n \in n_A^*(\varepsilon, x)$, then every $m \geq n$ also belongs to $n_A^*(\varepsilon, x)$. Since any intersection of upward-closed subsets of $\Nat$ is upward-closed, so is $n_A^*(\varepsilon)$.
    \item Let $\varepsilon' \ge \varepsilon$. If $\mathbb{P}_x(\tau_{A\cup\widetilde A} > n) \le \varepsilon$, then $\mathbb{P}_x(\tau_{A\cup\widetilde A} > n) \le \varepsilon'$, which gives the first inclusion. The second one follows by intersecting over $x \in S$.
\end{enumerate}

\end{proof}

In particular, whenever they are non-empty, the minima of these sets are non-increasing functions of $\varepsilon$.
The truncation sets are all non-empty, if and only if, the chain is decisive.  

\begin{restatable}{proposition}{PropNStarDecisive}
\label{prop:n-star-decisive}
Let $(S, P)$ be a Markov chain and $A \subseteq S$. The chain is decisive with respect to $A$ if and only if, for any $(\varepsilon, x)\in (0,1)\times S$, $n_A^*(\varepsilon, x) \neq \emptyset$.
\end{restatable}

\begin{proof}
The chain is decisive with respect to $A$ if and only if for every $x \in S$,  $\lim_{n \to \infty} \mathbb{P}_x(\tau_{A\cup\widetilde A} > n) = 0$ if and only if $\forall x \in S, \forall\varepsilon\in (0,1), \exists n \ge 0 : \mathbb{P}_x(\tau_{A\cup\widetilde A} > n) \le \varepsilon$ if and only if $\forall (\varepsilon, x)\in (0,1)\times S, n_A^*(\varepsilon, x) \neq \emptyset$.
\end{proof}

In contrast, the global set $n_A^*(\varepsilon)$ may be empty even under decisiveness as we will see later with Example~\ref{ex:pLCS-not-coarse}. The interest of truncation sets lies in the following approximation property.

\begin{restatable}{proposition}{PropDecToQuant}
\label{prop:dec-to-quant}
Let $(S, P)$ be a Markov chain, $A \subseteq S$ a target, $\varepsilon \in (0,1)$ and $x \in S$. Assume that $n_A^*(\varepsilon, x) \neq \emptyset$ and let $k \in n_A^*(\varepsilon, x)$. Then
$
  \mathbb{P}_x(\tau_A \leq k) \leq \mathbb{P}_x(\tau_A < \infty) \leq \mathbb{P}_x(\tau_A \leq k) + \varepsilon.
$
\end{restatable}
\begin{proof}
The first inequality is immediate. For the second one, we partition the event $\{\tau_A < \infty\}$ according to whether $A$ is reached within $k$ steps or strictly after $k$ steps:
$$
  \mathbb{P}_x(\tau_A < \infty) = \mathbb{P}_x(\tau_A \leq k) + \mathbb{P}_x(k < \tau_A < \infty).
$$
If a run reaches $A$ strictly after $k$ steps, then none of the states $X_0, \dots, X_k$ belongs to $A$, and none belongs to $\widetilde{A}$ either, since $A$ is not reachable from $\widetilde{A}$. Hence $\{k < \tau_A < \infty\} \subseteq \{\tau_{A \cup \widetilde{A}} > k\}$, and
$$
  \mathbb{P}_x(k < \tau_A < \infty) \leq \mathbb{P}_x(\tau_{A \cup \widetilde{A}} > k) \leq \varepsilon,
$$
where the last inequality follows from $k \in n_A^*(\varepsilon, x)$. Substituting this bound into the partition yields $\mathbb{P}_x(\tau_A < \infty) \leq \mathbb{P}_x(\tau_A \leq k) + \varepsilon$.
\end{proof}

 Combining Proposition~\ref{prop:dec-to-quant} with $\mathcal{A}_2$, any element of a truncation set reduces the infinite-horizon setting to a bounded-horizon one.

\begin{restatable}{corollary}{CorQuantFromHorizon}
\label{prop:quant-from-horizon}
Let $(\mathcal{S}, P)_{\mathrm{eff}}$ be an effective pWSTS, $B \subseteq S$ finite, $x \in S$, $\varepsilon \in \mathbb{Q} \cap (0,1)$ and $k \in n^*_{{\uparrow}B}(\varepsilon/2, x)$. Then $\mathcal{A}_2$, with horizon $k$ and precision $\varepsilon/2$, outputs a value $p^\ast \in \mathbb{Q}$ such that
$
  p^\ast \leq \mathbb{P}_x(\tau_{{\uparrow}B} < \infty) \leq p^\ast + \varepsilon.
$
\end{restatable}

\begin{proof}
By~\cite{fffy2026}, $p^\ast \le \mathbb{P}_x(\tau_{{\uparrow}B} \le k) \le p^\ast + \varepsilon/2$. By Proposition~\ref{prop:dec-to-quant} applied with $\varepsilon/2$, $\mathbb{P}_x(\tau_{{\uparrow}B} \le k) \le \mathbb{P}_x(\tau_{{\uparrow}B} < \infty) \le \mathbb{P}_x(\tau_{{\uparrow}B} \le k) + \varepsilon/2$. Combining both bounds gives the result.
\end{proof}

\subsection{Computing an element of $n^*_{{\uparrow}B}$}

We now turn to the computation of an element of $n^*_{{\uparrow}B}(\varepsilon, x)$ for an effective pWSTS $(\mathcal{S}, P)_{\mathrm{eff}}$ and a finite set $B \subseteq S$ such that $(\mathcal{S}, P)_{\mathrm{eff}}$ is decisive with respect to ${\uparrow}B$. Given $(\varepsilon, x) \in (0,1) \times S$, we write
$g_k = \mathbb{P}_x(\tau_{{\uparrow}B \cup \widetilde{{\uparrow}B}} \leq k)$,
so that $k \in n^*_{{\uparrow}B}(\varepsilon, x) \iff g_k \geq 1 - \varepsilon$. By~\cite{fffy2026}, the variant of $\mathcal{A}_2$ for the target ${\uparrow}B \cup \widetilde{{\uparrow}B}$ computes, for every horizon $k$ and every precision $\nu$, $\hat g_k \in \mathbb{Q}$ such that $\hat g_k \le g_k \le \hat g_k + \nu$. It thus suffices to enumerate $k = 0, 1, 2, \dots$, computing at each step such a lower approximation of $g_k$, until it exceeds $1 - \varepsilon$. Any value at which the scan stops belongs to $n^*_{{\uparrow}B}(\varepsilon, x)$, since $\hat g_k \le g_k$. Letting the precision $\nu_k$ depend on $k$, with $\lim_{k\to\infty}\nu_k=0$, ensures
termination: by decisiveness, $\lim_{k\to\infty}g_k=1$, so that the lower
approximations eventually exceed $1 - \varepsilon$. This yields
Procedure~\ref{alg:horizon-element}.

Let $B'$ be a finite basis of $\Pred^*({\uparrow}B)$, computed by the backward coverability algorithm. We define the function $\mathrm{In}$ which, given $y \in S$, decides whether $y \in {\uparrow}B$ and whether $y \in \Pred^*({\uparrow}B)$. Since $\le$ is decidable and $B$ and $B'$ are finite, $\mathrm{In}$ is computable: it suffices to test whether $b \le y$ for some $b \in B$, and whether $b' \le y$ for some $b' \in B'$.

\floatname{algorithm}{Procedure}
\begin{algorithm}[!htbp]
\caption{Computing an element of $n^*_{{\uparrow}B}(\varepsilon, x)$}
\label{alg:horizon-element}
\begin{algorithmic}[1]
\Require $B$, $B'$, $\mathrm{In}$, $x \in S$,
$\varepsilon \in \mathbb{Q} \cap (0,1)$, $\mathrm{succ}$, $P$, $\mathcal{A}_2$
\Ensure $k \in n^*_{{\uparrow}B}(\varepsilon, x)$
\State $k \gets 0$; \; $\hat g \gets 0$
\While{$\hat g \leq 1 - \varepsilon$}
  \State $\nu \gets 1/(k+2)$
  \State $\hat g \gets \mathcal{A}_2
  (B, B', \mathrm{In}, x, k, \nu, \mathrm{succ}, P)$
    \Comment{$\hat g \leq \mathbb{P}_x(\tau_{{\uparrow}B \cup \widetilde{{\uparrow}B}} \leq k)
    \leq \hat g + \nu$}
  \If{$\hat g \leq 1 - \varepsilon$} \State $k \gets k + 1$ \EndIf
\EndWhile
\State \Return $k$
\end{algorithmic}
\end{algorithm}

\begin{restatable}{lemma}{LemHorizonElementTermination}
\label{lem:horizon-element-termination}
If $(\mathcal{S}, P)_{\mathrm{eff}}$ is decisive with respect to ${\uparrow}B$, then Procedure~\ref{alg:horizon-element} terminates.
\end{restatable}
 \begin{proof}

Each iteration of the loop terminates, since the call at line 4 is to the variant of $\mathcal{A}_2$ for the target ${\uparrow}B \cup \widetilde{{\uparrow}B}$, which terminates for every finite horizon~\cite{fffy2026}. It remains to prove that the loop is exited after finitely many iterations. By decisiveness, $\mathbb{P}_x(\tau_{{\uparrow}B \cup \widetilde{{\uparrow}B}} < \infty) = 1$, hence $g_k \to 1$ as $k \to \infty$. At iteration $k$, the call returns a value $\hat g_k$ such that $\hat g_k \leq g_k \leq \hat g_k + \nu$, with $\nu = 1/(k+2)$, so that

$$
  \hat g_k \;\geq\; g_k - \frac{1}{k+2} .
$$

Since the right-hand side tends to $1$ and $1 > 1 - \varepsilon$, there exists $k_0 \in \Nat$ such that $g_{k_0} - 1/(k_0+2) > 1 - \varepsilon$. The exit condition is therefore satisfied at iteration $k_0$ at the latest.

\end{proof} 

Since termination is guaranteed under certain conditions, this procedure is now considered as a semi-algorithm.

\begin{restatable}{proposition}{PropHorizonElement}
\label{prop:horizon-element}
If $(\mathcal{S}, P)_{\mathrm{eff}}$ is decisive with respect to ${\uparrow}B$, then Semi-algorithm~\ref{alg:horizon-element}, combined with the backward coverability algorithm, computes an element of $n^*_{{\uparrow}B}(\varepsilon, x)$.
\end{restatable}

\begin{proof}
Termination is given by Lemma~\ref{lem:horizon-element-termination}. Let $k$ be the returned value and $\hat g_k$ the last computed value. The loop is exited because $\hat g_k > 1 - \varepsilon$. Since $\hat g_k$ is a lower bound on $g_k$, we get $g_k > 1 - \varepsilon$, that is, $\mathbb{P}_x(\tau_{{\uparrow}B \cup \widetilde{{\uparrow}B}} > k) < \varepsilon$. Hence $k \in n^*_{{\uparrow}B}(\varepsilon, x)$.
\end{proof}

\begin{remark}
\label{rem:horizon-element-decisive}
Semi-algorithm~\ref{alg:horizon-element} terminates on every pair $(\varepsilon, x) \in (\mathbb{Q} \cap (0,1)) \times S$ if and only if $(\mathcal{S}, P)_{\mathrm{eff}}$ is decisive with respect to ${\uparrow}B$. Lemma~\ref{lem:horizon-element-termination} proves one direction. Conversely, for a pair $(\varepsilon, x) \in (\mathbb{Q} \cap (0,1)) \times S$, if the semi-algorithm terminates, then since the correctness argument of Proposition~\ref{prop:horizon-element} does not rely on decisiveness, it returns an element of $n^*_{{\uparrow}B}(\varepsilon, x)$, which is therefore non-empty. Therefore, if it terminates on every pair, then $n^*_{{\uparrow}B}(\varepsilon, x) \neq \emptyset$ for every rational $\varepsilon \in (0,1)$ and every $x \in S$, hence for every real $\varepsilon \in (0,1)$ by Proposition~\ref{prop:n-star-properties}, since $n^*_{{\uparrow}B}(\varepsilon', x) \subseteq n^*_{{\uparrow}B}(\varepsilon, x)$ for any rationals $\varepsilon'$ such that $0 < \varepsilon' \le \varepsilon$. Proposition~\ref{prop:n-star-decisive} then yields decisiveness with respect to ${\uparrow}B$.
\end{remark}

\begin{remark}
\label{rem:proc3}
An element of $n^*_{{\uparrow}B}(\varepsilon, x)$ can also be extracted from $\mathcal{SA}_3$~\cite{fffy2026}. When run from $x$ with precision $\varepsilon$, $\mathcal{SA}_3$ exits its main loop at some depth $K \geq 1$ once the mass of its frontier is at most $\varepsilon/2$, while the mass discarded along the exploration is at most $\varepsilon/2$ as well. Together, these two bounds give $\mathbb{P}_x(\tau_{{\uparrow}B \cup \widetilde{{\uparrow}B}} > K - 1) \le \varepsilon$, that is, $K - 1 \in n^*_{{\uparrow}B}(\varepsilon, x)$ (see Appendix~\ref{annexe-proof-n-star-remark}).
\end{remark}

If $n^*_{{\uparrow}B}(\varepsilon, x)$ is non-empty, then it has a least element $m$. Semi-algorithm~\ref{alg:horizon-minimum} first computes an element $N \in n^*_{{\uparrow}B}(\varepsilon, x)$ with Semi-algorithm~\ref{alg:horizon-element}, and then computes $m$ by descending from $N$. This yields Procedure~\ref{alg:horizon-minimum}. Note that comparing $g_k$ with the threshold $1 - \varepsilon$ is not decidable in general: only the strict inequalities $g_k > 1 - \varepsilon$ and $g_k < 1 - \varepsilon$ can be certified, by refining the precision until the approximation settles the comparison. The descent tests $g_k$ for $k$ from $N - 1$ down to $\max(m - 1, 0)$. If $m = 0$, the descent reaches $k = 0$ and the loop guard $k > 0$ stops it, so no index below $m$ is tested. If $m > 0$, the last test concerns $k = m - 1$, and since $g_{m-1} < 1 - \varepsilon$ by minimality of $m$, this test always succeeds. In both cases, the semi-algorithm terminates and returns $\min n^*_{{\uparrow}B}(\varepsilon, x)$ provided $g_k \neq 1 - \varepsilon$ for every $k$ with $m \leq k < N$. 

Under the same assumption, the linear descent may be replaced by a binary search on $[0, N]$, which reduces the number of calls to $\mathcal{A}_2$ to $O(\log N)$. Indeed, every $k < m$ satisfies $g_k < 1 - \varepsilon$, so all the tests performed by the binary search terminate.

\begin{algorithm}[!htbp]
\caption{Computing $\min n^*_{{\uparrow}B}(\varepsilon, x)$}
\label{alg:horizon-minimum}
\begin{algorithmic}[1]
\Require $B$, $B'$, $\mathrm{In}$, $x \in S$,
$\varepsilon \in \mathbb{Q} \cap (0,1)$, $\mathrm{succ}$, $P$, $\mathcal{A}_2$
\Ensure $\min n^*_{{\uparrow}B}(\varepsilon, x)$
\State $N \gets$ output of Semi-algorithm~\ref{alg:horizon-element} on $(B, B', \mathrm{In}, x, \varepsilon, \mathrm{succ}, P, \mathcal{A}_2)$
  \Comment{$N \in n^*_{{\uparrow}B}(\varepsilon, x)$}
\State $k \gets N$; \; $\mathit{found} \gets \mathbf{false}$
  \Comment{invariant: $k \in n^*_{{\uparrow}B}(\varepsilon, x)$}
\While{$k > 0$ \textbf{and} $\neg\,\mathit{found}$}
  \State $\nu \gets 1$; \; $\mathit{settled} \gets \mathbf{false}$
  \While{$\neg\,\mathit{settled}$}
    \State $\hat g \gets \mathcal{A}_2
  (B, B', \mathrm{In}, x, k-1, \nu, \mathrm{succ}, P)$
    \If{$\hat g > 1 - \varepsilon$}
        \Comment{$k-1 \in n^*_{{\uparrow}B}(\varepsilon,x)$: descend}
      \State $k \gets k - 1$; \; $\mathit{settled} \gets \mathbf{true}$
    \ElsIf{$\hat g + \nu < 1 - \varepsilon$}
        \Comment{$k-1 \notin n^*_{{\uparrow}B}(\varepsilon,x)$: $k$ is the least}
      \State $\mathit{found} \gets \mathbf{true}$; \;
      $\mathit{settled} \gets \mathbf{true}$
    \Else
      \State $\nu \gets \nu / 2$ \Comment{inconclusive: refine}
    \EndIf
  \EndWhile
\EndWhile
\State \Return $k$
\end{algorithmic}
\end{algorithm}

\begin{restatable}{lemma}{LemHorizonMinimumTermination}
\label{lem:horizon-minimum-termination}
Let $m = \min n^*_{{\uparrow}B}(\varepsilon, x)$ and $N$ the value computed from Semi-algorithm~\ref{alg:horizon-element}. If $(\mathcal{S}, P)_{\mathrm{eff}}$ is decisive with respect to ${\uparrow}B$ and $g_k \neq 1 - \varepsilon$ for every $k$ with $m \leq k < N$, then Procedure~\ref{alg:horizon-minimum} terminates.
\end{restatable}
\begin{proof}
By Proposition~\ref{prop:horizon-element}, line 1 terminates and returns $N \in n^*_{{\uparrow}B}(\varepsilon, x)$. Each call at line 6 terminates~\cite{fffy2026} and returns a value $\hat g_{k-1}$ such that $\hat g_{k-1} \leq g_{k-1} \leq \hat g_{k-1} + \nu$.

Consider the inner loop for a given $k$. If $g_{k-1} > 1 - \varepsilon$, then the first test succeeds once $\nu$ is small enough, since $\hat g_{k-1} \geq g_{k-1} - \nu$. If $g_{k-1} < 1 - \varepsilon$, then $\hat g_{k-1} + \nu \leq g_{k-1} + \nu < 1 - \varepsilon$ once $\nu$ is small enough, and the second test succeeds. The only inconclusive case is $g_{k-1} = 1 - \varepsilon$. It is excluded by hypothesis when $k - 1 \geq m$, and it cannot occur when $k - 1 = m - 1$, since $g_{m-1} < 1 - \varepsilon$. Moreover, the semi-algorithm never tests a value $k - 1 < m - 1$, because it stops as soon as $k = m$. The inner loop therefore terminates after finitely many halvings of $\nu$.

The outer loop decreases $k$ by one at each iteration. It can only descend while $g_{k-1} > 1 - \varepsilon$, so $k$ never goes below $m$, and it stops at $k = m$ at the latest, since $g_{m-1} < 1 - \varepsilon$ triggers the second test (or $k = 0$ if $m = 0$). Hence it performs at most $N - m + 1$ iterations.
\end{proof}

Since termination is guaranteed under certain conditions, this procedure is now considered as a semi-algorithm.

\begin{restatable}{proposition}{PropHorizonMinimum}
\label{prop:horizon-minimum}
Under the assumptions of Lemma~\ref{lem:horizon-minimum-termination}, Semi-algorithm~\ref{alg:horizon-minimum}, combined with the backward coverability algorithm, returns $\min n^*_{{\uparrow}B}(\varepsilon, x)$.
\end{restatable}
\begin{proof}
We first show that the invariant $k \in n^*_{{\uparrow}B}(\varepsilon, x)$ holds throughout. It holds initially since $N \in n^*_{{\uparrow}B}(\varepsilon,x)$. The value of $k$ decreases only when $\hat g_{k-1} > 1 - \varepsilon$. Since $\hat g_{k-1} \leq g_{k-1}$, this gives $g_{k-1} > 1 - \varepsilon$, that is, $k - 1 \in n^*_{{\uparrow}B}(\varepsilon, x)$.

The loop exits in one of two ways. If $\mathit{found}$ becomes true, then $\hat g_{k-1} + \nu < 1 - \varepsilon$. Since $g_{k-1} \leq \hat g_{k-1} + \nu$, we get $g_{k-1} < 1 - \varepsilon$, so $k-1 \notin n^*_{{\uparrow}B}(\varepsilon,x)$. As $n^*_{{\uparrow}B}(\varepsilon,x)$ is upward-closed, no $j \leq k - 1$ belongs to it, and by the invariant $k$ is its least element. If the loop exits with $k = 0$, then $0 \in n^*_{{\uparrow}B}(\varepsilon, x)$ by the invariant, and $0$ is trivially the least element.
\end{proof}

\begin{remark}
\label{rem:exact-comparison}
Assume that $(\mathcal{S}, P)_{\mathrm{eff}}$ is finitely branching and
that its transition probabilities are rationals computed exactly, as is
the case for pVAS and pLCS (with $p_\ell \in \mathbb{Q}$). Then each
$g_k$ is a rational that can be computed exactly~\cite{fffy2026}, so its
comparison with the threshold $1 - \varepsilon$ is decidable. Hence, under
decisiveness alone, $\min n^*_{{\uparrow}B}(\varepsilon, x)$ is computable,
without the assumption $g_k \neq 1 - \varepsilon$.
\end{remark}

Note that, in the setting of Remark~\ref{rem:exact-comparison}, deciding membership in $n^*_{{\uparrow}B}(\varepsilon, x)$ and computing its least element are equivalent on non-empty instances: the least element yields a membership test by comparison, and conversely a membership test yields the least element by enumerating $k = 0, 1, 2, \dots$ until the first success.

\begin{remark}
\label{rem:cost}
Our interest, therefore, does not lie in simply computing an element of $n^*_{{\uparrow}B}(\varepsilon, x)$, but in computing one \emph{a priori}. The purpose of such an element is to give, before any exploration, information on the depth to be explored. Yet the semi-algorithms above, as well as $\mathcal{SA}_3$ in Remark~\ref{rem:proc3}, obtain it only by carrying out the exploration it was meant to delimit. Moreover, any pair $(\varepsilon', x')$ with $x' \neq x$ or $\varepsilon' < \varepsilon$ requires running the whole computation again, since no relation between the corresponding depths is available at this stage.
\end{remark}

In the sequel, we consider a target ${\uparrow}B$ and assume that the backward coverability algorithm has been applied once, which is required by the procedures of \cite{fffy2026}. We then seek to compute an element of $n^*_{{\uparrow}B}(\varepsilon, x)$ for any pair $(\varepsilon, x)$ using a number of elementary operations ($+, -, \times, /, \log, \dots$), whose dependence on $(\varepsilon, x)$ is as small as possible.

\subsection{Geometrically reachable sets}

In this subsection, we define a class of sets for which such a function can be computed. It builds on \emph{eagerness}, introduced by Abdulla, Ben Henda, Mayr and Sandberg~\cite{AbdullaBenHendaMayrSandbergATVA2006,AbdullaBenHendaMayrSandbergQEST2006,BenHenda08}.

\begin{definition}[Eagerness \cite{BenHenda08}]
Let $(S, P)$ be a Markov chain and $Q, F \subseteq S$. The chain is \emph{$Q$-eager} with respect to $F$ if there exist $\beta < 1$ and $C > 0$ such that for all $x \in Q$ and all $n \in \Nat$,
$
  \mathbb{P}_x\big(n \le \tau_F < \infty\big) \le C\beta^n .
$
\end{definition}

For $Q = \{x\}$, this is equivalent to the original definition of~\cite{AbdullaBenHendaMayrSandbergATVA2006}, which requires the bound $\beta_x^n$ only beyond some rank $N_x$ (see Appendix~\ref{app:eagerness}). For finite $Q$, $Q$-eagerness amounts to eagerness from each $x \in Q$, by taking maxima of the constants. For infinite $Q$, it is strictly stronger, since the prefactors may be unbounded and the rates may tend to~$1$. A related notion concerns return times. Recall that a non-empty set $A \subseteq S$ is an \emph{attractor} if $\mathbb{P}_x(\tau_A < \infty) = 1$ for every $x \in S$. A finite attractor $A$ is \emph{eager}~\cite{BenHenda08} if there exist $\beta < 1$ and $C \geq 1$ such that $\mathbb{P}_x(n < \tau_A^+ < \infty) \le C\beta^n$ for all $x \in A$ and $n \geq 0$. The chain is then $A$-eager with respect to any target set~\cite{AbdullaBenHendaMayrSandbergATVA2006,BenHenda08}. 

However, these notions were designed for other purposes and are not tailored to the bounds we need, for three main reasons.
\begin{itemize}
  \item \emph{No almost-sure reachability.} Eagerness only bounds the time to reach $F$ \emph{when $F$ is eventually reached}, since the measured events are intersected with $\{\tau_F < \infty\}$. A chain may thus be eager with respect to $x$ while $\mathbb{P}_x(\tau_F = \infty) > 0$. We bound the event $\{\tau_F > n\}$ instead, to capture both the speed and the occurrence of attraction. In particular, when $F = A \cup \widetilde{A}$, such a bound already implies decisiveness with respect to $A$ (Proposition~\ref{prop:eager-decisive}), which then need not be established separately.
  \item \emph{Too much uniformity.} $Q$-eagerness~\cite{BenHenda08} requires both the rate and the prefactor to be uniform over $Q$. This might be too strong for $Q = S$: a pLCS is not, in general, $S$-eager with respect to an upward-closed set (see Example~\ref{ex:pLCS-not-coarse}). Yet it does exhibit eager behavior over the whole of $S$ as soon as the prefactor is allowed to depend on the initial state (see Section~\ref{sec:beyond-coarseness}). We therefore keep a state-independent rate $\beta$ but allow a state-dependent prefactor $C_x$.
  \item \emph{Return time, not hitting time.} Eager attractors bound the \emph{return} time $\tau_A^+$ to a finite set $A$ from states within $A$, not the \emph{hitting} time $\tau_A$ from an arbitrary state. They thus give no information on how fast a state outside $A$ is first attracted to $A$. 
\end{itemize}
\begin{definition}[Geometrically reachable set]
\label{def:exp-hitting-set}
Let $(S, P)$ be a Markov chain and $Q, F \subseteq S$ non-empty. The set
$F$ is \emph{geometrically reachable with respect to $Q$} if there exist
$C \colon Q \to \mathbb{R}_{>0}$ and $\beta \in (0,1)$ such that, for all
$x \in Q$ and $n \geq 0$,
$
  \mathbb{P}_x(\tau_F > n) \leq C(x)\, \beta^n.
$
If moreover $C$ can be chosen constant, $F$ is \emph{uniformly
geometrically reachable with respect to $Q$}. When $Q = S$, we omit
``with respect to $Q$''.
\end{definition}

\begin{remark}
\label{rem:grs-n-star}
Let $A \subseteq S$ and assume that $A \cup \widetilde{A}$ is
geometrically reachable, with $C$ and $\beta$ as in
Definition~\ref{def:exp-hitting-set}. Since
$C(x)\, \beta^n \le \varepsilon$ as soon as
$n \geq \log(C(x)/\varepsilon)/\log(1/\beta)$, we get, for every
$\varepsilon \in (0,1)$ and $x \in S$,
$
  \max\left(0, \left\lceil \tfrac{\log(C(x)/\varepsilon)}{\log(1/\beta)}
  \right\rceil\right) \in n^*_{A}(\varepsilon, x).
$
Once $\beta$ and $C(x)$ are known, this gives an explicit element of
$n^*_{A}(\varepsilon, x)$, without exploring the chain. If
$A \cup \widetilde{A}$ is uniformly geometrically reachable, $C$ being
constant, the same expression gives an element of $n^*_{A}(\varepsilon)$,
which is thus non-empty. In either case,
$n^*_A(\varepsilon, x) \neq \emptyset$ for every
$(\varepsilon, x) \in (0,1) \times S$, so
Proposition~\ref{prop:n-star-decisive} immediately yields the following
proposition.
\end{remark}

\begin{proposition}
\label{prop:eager-decisive}
Let $(S, P)$ be a Markov chain and $A \subseteq S$ non-empty. If
$A \cup \widetilde{A}$ is geometrically reachable, then $(S, P)$ is
decisive with respect to $A$.
\end{proposition}

The next section focuses on uniformly geometrically reachable sets, which
yield horizons independent of the initial state. The non-uniform case is
treated in Section~\ref{sec:beyond-coarseness}.
\section{Uniformly geometrically reachable sets}
\label{sec:coarseness}

\subsection{Characterizing uniform geometric reachability}

In this subsection, we characterize \emph{uniform geometric reachability} through the notion of \emph{bounded coarseness}~\cite{AbdullaBenHendaMayrSandbergATVA2006}, and relate it to the global truncation set $n^*_A(\varepsilon)$. 

\begin{definition}[Boundedly coarse \cite{AbdullaBenHendaMayrSandbergATVA2006}]
A Markov chain $(S, P)$ is \emph{boundedly coarse with parameter
$(\delta, \ell) \in (0,1]\times\mathbb{N}$ towards a target $A$} if for all
$x \in S$, either $x \notin \Pred^*(A)$ or $\mathbb{P}_x(\tau_A \leq \ell) \geq \delta$.
\end{definition}

Within our framework, for a given subset $A \subseteq S$, the considered
target set is often $A \cup \widetilde{A}$.
Consequently, the condition $x \notin \Pred^*(A \cup \widetilde{A})$ never
holds for any $x \in S$, since $\Pred^*(A \cup \widetilde{A}) = S$. Thus,
being boundedly coarse with parameter $(\delta, \ell)$ towards
$A \cup \widetilde{A}$ is equivalent to requiring
$\mathbb{P}_x(\tau_{A \cup \widetilde{A}} \leq \ell) \geq \delta$ for all
$x \in S$.

\begin{remark}
\label{rem:coarse-params}
If $(S,P)$ is boundedly coarse with parameters $(\delta, \ell)$ towards a
target, then it is boundedly coarse with parameters $(\delta', \ell')$
towards the same target for all $\delta' \in (0, \delta]$ and
$\ell' \geq \ell$. In the sequel, whenever $(S,P)$ is assumed to be
boundedly coarse with parameters $(\delta, \ell)$, we therefore assume
without loss of generality that $\delta < 1$ and $\ell \geq 1$.
\end{remark}

The following lemma shows that global truncation sets and bounded coarseness towards
$A \cup \widetilde{A}$ are two readings of the same condition.

\begin{restatable}{lemma}{LemNstarCoarse}
\label{lem:nstar-coarse}
Let $A \subseteq S$, $\varepsilon \in (0,1)$ and $\ell \in \mathbb{N}$. Then
$\ell \in n^*_A(\varepsilon)$ if and only if $(S,P)$ is boundedly coarse
with parameters $(1-\varepsilon, \ell)$ towards $A \cup \widetilde{A}$.
\end{restatable}
\begin{proof}
By definition, $\ell \in n^*_A(\varepsilon)$ if and only if
$\mathbb{P}_x(\tau_{A \cup \widetilde{A}} > \ell) \leq \varepsilon$ for all
$x \in S$, that is,
$\mathbb{P}_x(\tau_{A \cup \widetilde{A}} \leq \ell) \geq 1 - \varepsilon$
for all $x \in S$. Since $\Pred^*(A \cup \widetilde{A}) = S$, this is exactly
bounded coarseness with parameters $(1-\varepsilon, \ell)$ towards
$A \cup \widetilde{A}$.
\end{proof}

We now characterize uniform geometric reachability in terms of bounded
coarseness. The implication from bounded coarseness to eagerness was
established in~\cite[Lemma~5.1]{AbdullaBenHendaMayrSandbergATVA2006}, and
its proof adapts to uniform geometric reachability. 

\begin{restatable}{lemma}{coarseImpliesEager}
\label{lem:coarse-implies-eager}
If $(S, P)$ is boundedly coarse with parameters $(\delta,
\ell)$ towards $A \cup \widetilde{A}$, then $A \cup \widetilde{A}$ is a
uniformly geometrically reachable set.
\end{restatable}

\begin{proof}
This is Lemma~5.1 in~\cite{AbdullaBenHendaMayrSandbergATVA2006}, translated
to our setting by replacing the target set with $A \cup \widetilde{A}$. For
the sake of completeness and because it is useful for what follows, we
provide the full proof below. By Remark~\ref{rem:coarse-params}, we assume
$\delta < 1$ and $\ell \geq 1$. Since
$\Pred^*(A \cup \widetilde{A}) = S$, bounded coarseness gives
$\mathbb{P}_y(\tau_{A \cup \widetilde{A}} > \ell) \leq 1 - \delta$ for every
$y \in S$.

We first observe that, for every $m \geq 0$,
$$
  \mathbf{1}_{\{\tau_{A \cup \widetilde{A}} > (m+1)\ell\}}
  = \mathbf{1}_{\{\tau_{A \cup \widetilde{A}} > m\ell\}}
  \cdot \big(\mathbf{1}_{\{\tau_{A \cup \widetilde{A}} > \ell\}} \circ \theta_{m\ell}\big).
$$
Indeed, $\tau_{A \cup \widetilde{A}} > (m+1)\ell$ holds if and only if
$X_k \notin A \cup \widetilde{A}$ for every $k \leq (m+1)\ell$, that is, if
and only if $X_k \notin A \cup \widetilde{A}$ for every $k \leq m\ell$ and
$X_j \circ \theta_{m\ell} = X_{m\ell + j} \notin A \cup \widetilde{A}$ for
every $j \leq \ell$. The latter condition is
$\tau_{A \cup \widetilde{A}} \circ \theta_{m\ell} > \ell$.

We show by induction on $m$ that
$\mathbb{P}_x(\tau_{A \cup \widetilde{A}} > m\ell) \leq (1-\delta)^m$ for
every $x \in S$ and $m \geq 0$. The base case $m=0$ is immediate. Assume the
bound holds at $m$ and let $x \in S$. Since $\tau_{A \cup \widetilde{A}}$ is
a stopping time, $\{\tau_{A \cup \widetilde{A}} > m\ell\} \in
\mathcal{F}_{m\ell}$. Conditioning on $\mathcal{F}_{m\ell}$ and applying
Proposition~\ref{prop:markov} with
$\Phi = \mathbf{1}_{\{\tau_{A \cup \widetilde{A}} > \ell\}}$ and $n = m\ell$,
we obtain
\begin{align*}
  \mathbb{P}_x(\tau_{A \cup \widetilde{A}} > (m+1)\ell)
  &= \mathbb{E}_x\!\left[\mathbf{1}_{\{\tau_{A \cup \widetilde{A}} > m\ell\}}
    \cdot \big(\mathbf{1}_{\{\tau_{A \cup \widetilde{A}} > \ell\}} \circ \theta_{m\ell}\big)\right] \\
  &= \mathbb{E}_x\!\left[\mathbf{1}_{\{\tau_{A \cup \widetilde{A}} > m\ell\}}
    \, \mathbb{E}_x\!\left[\mathbf{1}_{\{\tau_{A \cup \widetilde{A}} > \ell\}} \circ \theta_{m\ell}
    \,\middle|\, \mathcal{F}_{m\ell}\right]\right] \\
  &= \mathbb{E}_x\!\left[\mathbf{1}_{\{\tau_{A \cup \widetilde{A}} > m\ell\}}
    \, \mathbb{P}_{X_{m\ell}}(\tau_{A \cup \widetilde{A}} > \ell)\right] \\
  &\leq (1-\delta)\, \mathbb{P}_x(\tau_{A \cup \widetilde{A}} > m\ell) \\
  &\leq (1-\delta)^{m+1}.
\end{align*}
The first inequality uses
$\mathbb{P}_y(\tau_{A \cup \widetilde{A}} > \ell) \leq 1 - \delta$ for every
$y \in S$, and the second the induction hypothesis.

For general $n \geq 0$, write $n = m\ell + r$ with $0 \leq r < \ell$. Since
$\{\tau_{A \cup \widetilde{A}} > n\} \subseteq
\{\tau_{A \cup \widetilde{A}} > m\ell\}$,
$$
  \mathbb{P}_x(\tau_{A \cup \widetilde{A}} > n)
  \leq (1-\delta)^m
  = (1-\delta)^{(n-r)/\ell}
  \leq (1-\delta)^{-1} \big[(1-\delta)^{1/\ell}\big]^n.
$$
Setting $\beta = (1-\delta)^{1/\ell} \in (0,1)$ and $C = (1-\delta)^{-1} > 0$,
both independent of $x$, gives
$\mathbb{P}_x(\tau_{A \cup \widetilde{A}} > n) \leq C\beta^n$ for every
$x \in S$ and $n \geq 0$: $A \cup \widetilde{A}$ is a uniformly
geometrically reachable set.
\end{proof}

Uniform geometric reachability of $A \cup \widetilde{A}$, bounded coarseness towards $A \cup \widetilde{A}$, and the non-emptiness of the global truncation sets are in fact equivalent. Consequently, the converse of Lemma~\ref{lem:coarse-implies-eager} also
holds, which has no counterpart for eagerness.

\begin{restatable}{theorem}{ThmUgrCharacterization}
\label{thm:ugr-characterization}
Let $A \subseteq S$. The following statements are equivalent:
\begin{enumerate}
  \item $(S,P)$ is boundedly coarse towards $A \cup \widetilde{A}$ for some parameters $(\delta,\ell)$;
  \item $A \cup \widetilde{A}$ is a uniformly geometrically reachable set;
  \item $n^*_A(\varepsilon) \neq \emptyset$ for every $\varepsilon \in (0,1)$;
  \item $n^*_A(\varepsilon_0) \neq \emptyset$ for some $\varepsilon_0 \in (0,1)$.
\end{enumerate}
\end{restatable}


\begin{proof}
We prove the cycle of implications $(1) \Rightarrow (2) \Rightarrow (3) \Rightarrow (4) \Rightarrow (1)$.
\begin{enumerate}
  \item []  $(1) \Rightarrow (2)$: This is Lemma~\ref{lem:coarse-implies-eager}.
  
  \item[] $(2) \Rightarrow (3)$: Let $C > 0$ and $\beta \in (0,1)$ be such that $\mathbb{P}_x(\tau_{A \cup \widetilde{A}} > n) \le C \beta^n$ for all $x \in S$ and $n \in \mathbb{N}$, and let $\varepsilon \in (0,1)$. For every $n \geq \max(0,\log(C/\varepsilon)/\log(1/\beta))$, we have $C\beta^n \le \varepsilon$, hence $\mathbb{P}_x(\tau_{A \cup \widetilde{A}} > n) \le \varepsilon$ for every $x \in S$, that is, $n \in n^*_A(\varepsilon)$. In particular, $n^*_A(\varepsilon) \neq \emptyset$.
  
  \item[] $(3) \Rightarrow (4)$:  Immediate.
  
  \item[] $(4) \Rightarrow (1)$:  This is Lemma~\ref{lem:nstar-coarse}, with parameters $(1-\varepsilon_0, \ell)$ for any $\ell \in n^*_A(\varepsilon_0)$.
\end{enumerate}
\end{proof}

\begin{example}[A pLCS without uniform geometric reachability]
\label{ex:pLCS-not-coarse}
Consider a pLCS with a single channel, $Q = \{q_0, q_1\}$,
$\Sigma = \{a, b\}$,
$\Delta = \{(q_0, !b, q_0),\ (q_0, ?b, q_1),\ (q_1, \tau, q_1)\}$,
$W$ constant equal to $1$ and $p_\ell \in (0,1) \cap \mathbb{Q}$, and let
$A = {\uparrow}\{(q_1, \epsilon)\}$. Every state of $S$ can reach $A$  by
losing all its messages, sending $b$ and receiving it, so
$\widetilde{A} = \emptyset$ and $S = \Pred^*(A)$. Let $x_L = (q_0, a^L b)$ and $n \geq 0$. Since $\widetilde{A} = \emptyset$,
$\tau_{A \cup \widetilde{A}} = \tau_A$ is the first visit to $q_1$. This
visit requires firing $(q_0, ?b, q_1)$, hence having $b$ at the head of the
channel after a loss phase. No rule sends or receives $a$, and $!b$ appends
$b$ at the tail, so the channel content stays in $a^* b^*$: $b$ can be at
the head only once the $L$ occurrences of $a$ have all been lost. Each of
them remains in the channel until it is lost, and at each step it is lost
with probability $p_\ell$, independently of the others. It is thus lost
within $n$ steps with probability $1 - (1 - p_\ell)^n$, independently of
the others, and
$
  \mathbb{P}_{x_L}(\tau_{A \cup \widetilde{A}} \leq n)
  \leq \big(1 - (1 - p_\ell)^n\big)^L
  \xrightarrow[L \to \infty]{} 0 .
$
Thus, this illustrates two related points. First, in pLCS, the union of an upward-closed set and its unreachability zone is not, in general, uniformly geometrically reachable. This follows from the converse of Lemma~\ref{lem:coarse-implies-eager}, established by Theorem~\ref{thm:ugr-characterization}. Moreover, since pLCS are decisive with respect to upward-closed sets~\cite{AbdullaBenHendaMayr2007} and $\widetilde{A} = \emptyset$, $A$ is thus reached almost surely, so that $\mathbb{P}_{x_L}(n \le \tau_A < \infty) = \mathbb{P}_{x_L}(\tau_A \geq n)$. Consequently, pLCS are generally not $S$-eager with respect to such unions.  However, we will see in Section~\ref{sec:beyond-coarseness} that such unions are geometrically reachable.
\end{example}

\subsection{\emph{A priori} global horizon}

\begin{restatable}{proposition}{PropBootstrap}
\label{prop:bootstrap}
Let $A \subseteq S$, $\varepsilon_0 \in (0,1)$ and $\ell \in n^*_A(\varepsilon_0)$. Then, for every $\varepsilon \in (0,1)$,
$
  \ell \left\lceil \frac{\log \varepsilon}{\log \varepsilon_0} \right\rceil \in n^*_A(\varepsilon).
$
\end{restatable}

\begin{proof}
Let $F = A \cup \widetilde{A}$ and $x \in S$. For every $k \geq 0$, we have $\{\tau_F > (k+1)\ell\} = \{\tau_F > k\ell\} \cap \{\tau_F \circ \theta_{k\ell} > \ell\}$. Since $\{\tau_F > k\ell\} \in \mathcal{F}_{k\ell}$, the Markov property (Proposition~\ref{prop:markov}) gives
$$
  \mathbb{P}_x(\tau_F > (k+1)\ell)
  = \mathbb{E}_x\big[\mathbf{1}_{\{\tau_F > k\ell\}}\, \mathbb{P}_{X_{k\ell}}(\tau_F > \ell)\big]
  \le \varepsilon_0\, \mathbb{P}_x(\tau_F > k\ell),
$$
where the inequality uses $\ell \in n^*_A(\varepsilon_0)$, that is, $\mathbb{P}_y(\tau_F > \ell) \le \varepsilon_0$ for every $y \in S$. By induction, $\mathbb{P}_x(\tau_F > k\ell) \le \varepsilon_0^k$ for every $k \geq 0$. Let $\varepsilon \in (0,1)$ and $k = \lceil \log \varepsilon / \log \varepsilon_0 \rceil$. Then $\varepsilon_0^k \le \varepsilon$, so that $\mathbb{P}_x(\tau_F > k\ell) \le \varepsilon$ for every $x \in S$, that is, $k\ell \in n^*_A(\varepsilon)$.
\end{proof}

By Theorem~\ref{thm:ugr-characterization}, bounded coarseness towards $A \cup \widetilde{A}$ amounts to the existence of an element of a global truncation set. By Lemma~\ref{lem:nstar-coarse}, bounded coarseness with parameters
$(\delta, \ell)$ towards $A \cup \widetilde{A}$ gives
$\ell \in n^*_A(1-\delta)$, so Proposition~\ref{prop:bootstrap} yields
horizons for every precision directly from these parameters. In practice, bounded coarseness is often established towards the target $A$
itself. It transfers to $A \cup \widetilde{A}$, so that the horizons above
apply in that case as well. 

\begin{restatable}{lemma}{LemCoarseTransfer}
\label{lem:coarse-transfer}
Let $A \subseteq S$. If $(S,P)$ is boundedly coarse with parameters $(\delta, \ell)$ towards $A$, then it is boundedly coarse with the same parameters towards $A \cup \widetilde{A}$.
\end{restatable}

\begin{proof}
Let $x \in S$. If $x \in \Pred^*(A)$, then $\mathbb{P}_x(\tau_A \leq \ell) \geq \delta$ by assumption, and $\tau_{A \cup \widetilde{A}} \leq \tau_A$ gives $\mathbb{P}_x(\tau_{A \cup \widetilde{A}} \leq \ell) \geq \delta$. Otherwise, $x \in \widetilde{A}$ by definition, so $\tau_{A \cup \widetilde{A}} = 0$ and the probability equals $1$. In both cases $\mathbb{P}_x(\tau_{A \cup \widetilde{A}} \leq \ell) \geq \delta$, and the exclusion clause of the definition is never needed since $\Pred^*(A \cup \widetilde{A}) = S$.
\end{proof}

\begin{example}
The converse of Lemma~\ref{lem:coarse-transfer} does not hold. Let $S = \mathbb{N} \cup \{a, z\}$, where $a$ and $z$ are absorbing and each $n \in \mathbb{N}$ moves to $a$ with probability $2^{-n}$ and to $z$ with probability $1 - 2^{-n}$. Take $A = \{a\}$, so that $\widetilde{A} = \{z\}$. The chain is boundedly coarse with parameters $(1,1)$ towards $A \cup \widetilde{A}$, but $\mathbb{P}_n(\tau_A \leq K) = 2^{-n}$ for every $K \geq 1$, so it is not boundedly coarse towards $A$. 
\end{example}

It therefore allows us to establish the following theorem in the case of
bounded coarseness.

\begin{restatable}{theorem}{ThmComputableHorizon}
\label{thm:computable-horizon}
If $(S, P)$ is boundedly coarse towards $A$ with given parameters
$(\delta, \ell) \in ((0,1] \cap \mathbb{Q}) \times \mathbb{N}$, then there exists $\mathcal{N}_A \colon \mathbb{Q} \cap (0,1) \to \Nat$ such
that:
\begin{itemize}
  \item the evaluation of $\mathcal{N}_A(\varepsilon)$ requires a constant (independent of
  $\varepsilon$)
  number of elementary operations $(-, \times, /, \log)$ followed by a rational upper approximation and a ceiling;
  \item $\mathcal{N}_A(\varepsilon) \in n^*_A(\varepsilon)$;
  \item $\mathcal{N}_A(\varepsilon) = O\big(\log(1/\varepsilon)\big)$ as
  $\varepsilon \to 0$.
\end{itemize}
\end{restatable}
\begin{proof}
By Lemma~\ref{lem:coarse-transfer}, $(S,P)$ is boundedly coarse with
parameters $(\delta, \ell)$ towards $A \cup \widetilde{A}$, and we assume
$\delta \in (0,1)$ and $\ell \geq 1$ (see Remark~\ref{rem:coarse-params}).
For $\varepsilon \in \mathbb{Q} \cap (0,1)$, let $q(\varepsilon) \in \mathbb{Q}$ be such that
$0 \leq q(\varepsilon) - \log \varepsilon / \log(1-\delta) \leq 1$, and set
$
  \mathcal{N}_A(\varepsilon) = \ell \, \lceil q(\varepsilon) \rceil .
$
By Lemma~\ref{lem:nstar-coarse}, $\ell \in n^*_A(1-\delta)$, and the proof of
Proposition~\ref{prop:bootstrap} with $\varepsilon_0 = 1-\delta$ shows that
$k\ell \in n^*_A(\varepsilon)$ for every integer
$k \geq \log \varepsilon / \log(1-\delta)$; in particular
$\mathcal{N}_A(\varepsilon) \in n^*_A(\varepsilon)$. Since $\delta$ and
$\ell$ are fixed, the evaluation of $\mathcal{N}_A(\varepsilon)$ requires a constant
  number of elementary operations $(-, \times, /, \log)$, followed by a rational upper approximation and a ceiling; and
$\mathcal{N}_A(\varepsilon) \leq \ell\,\big(\log \varepsilon /
\log(1-\delta) + 2\big) = O(\log(1/\varepsilon))$.
\end{proof}

\begin{remark}
\label{rem:min-log}
Since $\mathcal{N}_A(\varepsilon) \in n^*_A(\varepsilon)$,
$\min n^*_A(\varepsilon)$ is itself a $O\big(\log(1/\varepsilon)\big)$ as
$\varepsilon \to 0$. Moreover, since
$n^*_A(\varepsilon) \subseteq n^*_A(\varepsilon, x)$ for every $x \in S$,
this bound holds uniformly on the least elements of the sets
$n^*_A(\varepsilon, x)$.
\end{remark}

\subsection{$\gamma$-coarse pWSTS}

Recall that a Markov chain is \emph{$\gamma$-coarse} if every
positive transition probability is bounded below by $\gamma >
0$~\cite{AbdullaBenHendaMayrSandbergATVA2006}.
Let us recall Lemma~5.2 from \cite{AbdullaBenHendaMayrSandbergATVA2006}.

\begin{lemma}[Lemma~5.2 \cite{AbdullaBenHendaMayrSandbergATVA2006}]
\label{lemma:5.2}
Let $(S, P)$ be a Markov chain and let $A \subseteq S$. If it
is $\gamma$-coarse and there exists $\ell_0 \in \Nat$ such that
$\Pred^{\leq\ell_0}(A) = \Pred^*(A)$, then it is boundedly coarse with
parameter $(\gamma^{\ell_0}, \ell_0)$.
\end{lemma}

For effective pWSTS, the integer $\ell_0$ can be computed with the backward algorithm. Combining this computability with Lemma~\ref{lemma:5.2} yields the following proposition.

\begin{restatable}{proposition}{PropTheoremCoarse}
\label{theorem:coarse}
Let $(\mathcal{S},P)_{\mathrm{eff}}$ be an effective $\gamma$-coarse pWSTS with
$\gamma \in (0,1) \cap \mathbb{Q}$ and $B \subseteq S$ a finite set. Then a
function $\mathcal{N}_{\uparrow B}$ as in Theorem~\ref{thm:computable-horizon} is
computable.
\end{restatable}

\begin{proof}
The backward coverability algorithm computes the least $\ell_0$ such that
$\Pred^{\le \ell_0}({\uparrow}B) = \Pred^*({\uparrow}B)$. By
Lemma~\ref{lemma:5.2}, $(\mathcal{S},P)$ is boundedly coarse with parameters
$(\gamma^{\ell_0}, \ell_0)$ towards ${\uparrow}B$, which can be taken with
$\gamma^{\ell_0} < 1$ and $\ell_0 \geq 1$ by Remark~\ref{rem:coarse-params}.
The proof of Theorem~\ref{thm:computable-horizon} then provides a function
$\mathcal{N}_{\uparrow B}$ with the required properties, namely
$\mathcal{N}_{\uparrow B}(\varepsilon) = \ell \lceil q(\varepsilon) \rceil$,
where $(\delta, \ell)$ are the above parameters and $q(\varepsilon) \in \mathbb{Q}$
satisfies $0 \leq q(\varepsilon) - \log \varepsilon / \log(1-\delta) \leq 1$.

It remains to show that this function is computable. First, $\ell_0$ is
computed by the backward coverability algorithm, and $\gamma \in \mathbb{Q}$,
so the parameters $\delta \in (0,1) \cap \mathbb{Q}$ and $\ell \in \mathbb{N}$
are computable. Now let $\varepsilon \in \mathbb{Q} \cap (0,1)$, and set
$x(\varepsilon) = \log \varepsilon / \log(1-\delta)$. As a quotient of
computable reals with nonzero denominator ($\log(1-\delta) < 0$),
$x(\varepsilon)$ is a computable real, and the map
$\varepsilon' \mapsto x(\varepsilon')$ is computable. Hence we can compute a
rational $r$ with $|r - x(\varepsilon)| \leq 1/2$. Then
$q(\varepsilon) = r + 1/2$ is a rational with
$0 \leq q(\varepsilon) - x(\varepsilon) \leq 1$. Finally, the ceiling of a
rational is computable. Therefore
$\mathcal{N}_{\uparrow B}(\varepsilon) = \ell \lceil q(\varepsilon) \rceil$ is
computable from $\varepsilon$.
\end{proof}
We now instantiate our framework with pVAS, which are both effective pWSTS \cite{fffy2026} and $\gamma$-coarse.

\begin{restatable}{corollary}{CorPvassEager}
\label{cor:pvass-eager}
Let $d \in \Nat_+$, $V \subseteq \mathbb{Z}^d$ be a VAS with weight function
$w \colon V \to \Nat_+$, and $B \subseteq \mathbb{N}^d$ a finite set. Then a
function $\mathcal{N}_{\uparrow B}$ as in Theorem~\ref{thm:computable-horizon} is
computable.
\end{restatable}

\begin{proof}
Let $W = \sum_{v \in V} w(v)$ and $w_{\min} = \min_{v \in V} w(v)$. Every
positive transition probability is at least $w_{\min}/W$, so the associated pVAS is a $\gamma$-coarse pWSTS with
$\gamma = \min(w_{\min}/W, 1/2) \in (0,1) \cap \mathbb{Q}$, and
Proposition~\ref{theorem:coarse} applies since pVAS are effective pWSTS.
\end{proof}

\begin{remark}
In a $\gamma$-coarse pWSTS such as pVAS, the branching degree is necessarily finite, as already noted
in~\cite{AbdullaBenHendaMayrSandbergATVA2006}. Indeed, let $x \in S$ be a non-deadlock state, the claim being trivial otherwise. Since $P$ is compatible with the transition relation, we have $\Post(x) = \mathrm{supp}(P(x))$ for non-deadlock states. Because all positive transition probabilities are bounded below by $\gamma$, we obtain
$
  1 = \sum_{y \in \Post(x)} P(x)(y) \geq |\Post(x)| \cdot \gamma,
$
hence $|\Post(x)| \leq b$, where $b = \lfloor 1/\gamma \rfloor$. 

This yields a worst-case bound on the size of the exploration. By Lemma~\ref{lem:nstar-coarse} and Proposition~\ref{prop:bootstrap}, applied
with $\varepsilon_0 = 1 - \gamma^{\ell_0}$ (with $\ell_0 \ge 1$) and the parameters
$(\gamma^{\ell_0}, \ell_0)$ of Proposition~\ref{theorem:coarse}, the horizon
$
  N = \ell_0 \left\lceil \frac{\log \varepsilon}{\log(1-\gamma^{\ell_0})} \right\rceil
$
belongs to $n^*_{{\uparrow}B}(\varepsilon, x)$ for every $x \in S$. Since at most $b^k$ states are reached at depth $k$, exploring all runs of length at most $N$ from $x$ visits at most
$
  \sum_{k=0}^{N} b^k \;\le\; 2\, b^{N}
$
states in total when $b \geq 2$, and at most $N + 1$ states when $b = 1$. 

The converse, however, is false: finite branching does not imply $\gamma$-coarseness. For instance, the chain on $\Nat$ in which each $n$ moves to $n+1$ with probability $2^{-(n+1)}$ and to $0$ otherwise is finitely branching, but its positive transition probabilities are not bounded below by any $\gamma > 0$.
\end{remark}

\subsection{Stochastically monotone pWSTS}

\medskip

\begin{definition}\cite{fffy2026}
\label{def:stochastic-monotonicity}
A pWSTS $(\mathcal{S}, P)$ is \emph{stochastically monotone} if for all
$x, x' \in S$ with $x \leq x'$ and every upward-closed set $U \subseteq S$,
$
  P(x)(U) \;\leq\; P(x')(U).
$
\end{definition}

From \cite{fffy2026}, recall that a stochastically monotone pWSTS is boundedly coarse towards any
non-empty upward-closed set $U$,
with parameters $(\delta, \ell)$ defined as follows. Let
$B' = \{b'_1, \dots, b'_r\}$ be a finite basis of $\Pred^*(U)$ and, for each
$1 \leq j \leq r$, let $k_j$ and $\delta_j$ be the length and the probability
of a finite run from $b'_j$ to $U$. Then
$
  \ell = \max\Big(1, \max_{1 \leq j \leq r} k_j\Big)
  $ and $
  \delta = \min_{1 \leq j \leq r} \delta_j > 0.
$ For an effective pWSTS, valid parameters $(\delta, \ell)$ can be computed
by combining the backward coverability algorithm with $\mathcal{A}_2$.
Alternatively, $\mathcal{A}_1$, which directly returns a
shortest run from each $b'_j$ to $U$ together with its probability, can
be used and might be faster in practice.

\begin{restatable}{proposition}{ThmComputableMonotone}
\label{thm:computable-monotone}
Let $(\mathcal{S}, P)_{\mathrm{eff}}$ be an effective pWSTS that is stochastically monotone and
$B \subseteq S$ a finite set. Then a function $\mathcal{N}_{\uparrow B}$ as in
Theorem~\ref{thm:computable-horizon} is computable.
\end{restatable}

\begin{proof}
If $B = \emptyset$, then $n^*_{\emptyset}(\varepsilon) = \mathbb{N}$ for every
$\varepsilon$, so any constant function is suitable. Otherwise, the parameters
$(\delta, \ell)$ recalled above are computable. We take the candidate function
$\mathcal{N}_{\uparrow B}(\varepsilon) = \ell \lceil q(\varepsilon) \rceil$
from the proof of Theorem~\ref{thm:computable-horizon}, and its computability
follows as in the proof of Proposition~\ref{theorem:coarse}.
\end{proof}

Also from \cite{fffy2026}, let us recall that multi-type Galton--Watson processes associated with a family of laws of reproduction $\mu = (\mu^{(i)})_{1 \leq i \leq d}$ are effective pWSTS provided that, for any $1 \leq i \leq d$: (i)~$\mu^{(i)}$ is computable, (ii)~membership in $\mathrm{supp}(\mu^{(i)})$ is decidable, and (iii)~determining, for any $v$, whether there exists $v' \in \mathrm{supp}(\mu^{(i)})$ such that $v' \ge v$ is decidable. They are furthermore stochastically monotone pWSTS \cite{fffy2026}.
\begin{corollary}
\label{cor:gw-computable}
Let $d \in \Nat_+$, $(Z_n)_{n \in \Nat}$ be a multi-type Galton--Watson
process associated with a family of laws of reproduction
$\mu = (\mu^{(i)})_{1 \leq i \leq d}$ satisfying the three conditions above,
and $B \subseteq \Nat^d$ a finite set. Then a function $\mathcal{N}_{\uparrow B}$ as in
Theorem~\ref{thm:computable-horizon} is computable.
\end{corollary}


\section{Beyond uniformity}
\label{sec:beyond-coarseness}

Although bounded coarseness leads to powerful conclusions, many Markov chains fail
to satisfy it, among them pLCS (see Example~\ref{ex:pLCS-not-coarse}). To
extend our quantitative framework to such systems, we start from the pLCS
case admitting an eager attractor, treated
in~\cite{AbdullaBenHendaMayrSandbergATVA2006,BenHenda08}, and extend it
to a subclass of Markov chains admitting a finite geometrically reachable set.
Then, building on the drift theory of Markov
chains~\cite{Hajek1982,meyn1993markov}, we introduce a
method to construct a finite geometrically reachable set.

\subsection{Markov chains with a finite geometrically reachable set}

\begin{restatable}{lemma}{LemEagernessTransitivity}
\label{thm:eagerness-transitivity}
Let $(S, P)$ be a Markov chain and $A, T \subseteq S$ non-empty. Assume
that $A$ is geometrically reachable, with
$C_A \colon S \to \mathbb{R}_{>0}$ and $\beta_A \in (0,1)$, and that
$T$ is uniformly geometrically reachable with respect to $A$, with
$C_T > 0$ and $\beta_T \in (0,1)$. Then $T$ is geometrically reachable,
with $C(x) = C_A(x)\, \beta_A^{-1/2} + C_T$ and
$\beta = \max(\beta_A, \beta_T)^{1/2}$.
\end{restatable}\begin{proof}
Fix $x \in S$, $n \geq 0$, and let $m = \lfloor n/2 \rfloor$. We decompose
the event $\{\tau_T > n\}$ according to whether $A$ is reached by time $m$,
which yields
$$
  \mathbb{P}_x(\tau_T > n) \leq
  \mathbb{P}_x(\tau_T > n,\, \tau_A > m) +
  \mathbb{P}_x(\tau_T > n,\, \tau_A \leq m).
$$
For the first term, bounding the joint probability by the marginal event
and using assumption (1) gives
$\mathbb{P}_x(\tau_T > n,\, \tau_A > m) \leq \mathbb{P}_x(\tau_A > m) \leq
C_A(x)\, \beta_A^m$.
For the second term, on the event $\{\tau_T > n\} \cap \{\tau_A \leq m\}$ we
have $\tau_A \leq m < \tau_T$. Thus $T$ is not visited before $\tau_A$, and
the remaining time to reach $T$ after passing through $A$ satisfies
$\tau_T \circ \theta_{\tau_A} = \tau_T - \tau_A > n - m = \lceil n/2
\rceil$. Since $\{\tau_A \leq m\} \in \mathcal{F}_{\tau_A}$ and
$X_{\tau_A} \in A$, the strong Markov property combined with assumption (2)
yields:
\begin{align*}
  \mathbb{P}_x(\tau_A \leq m,\, \tau_T \circ \theta_{\tau_A} > \lceil n/2 \rceil)
  &= \mathbb{E}_x \left[ \mathbf{1}_{\tau_A \leq m} \,
     \mathbb{E}_x \left[ \mathbf{1}_{\tau_T \circ \theta_{\tau_A} >
     \lceil n/2 \rceil} \mid \mathcal{F}_{\tau_A} \right] \right] \\
  &= \mathbb{E}_x \left[ \mathbf{1}_{\tau_A \leq m} \,
     \mathbb{P}_{X_{\tau_A}}(\tau_T > \lceil n/2 \rceil) \right] \\
  &\leq \mathbb{E}_x \left[ \mathbf{1}_{\tau_A \leq m} \right]
     \sup_{a \in A} \mathbb{P}_a(\tau_T > \lceil n/2 \rceil) \\
  &\leq \sup_{a \in A} \mathbb{P}_a(\tau_T > \lceil n/2 \rceil)
     \leq C_T\, \beta_T^{\lceil n/2 \rceil}.
\end{align*}
Summing both contributions and applying $\lfloor n/2 \rfloor \geq (n-1)/2$
and $\lceil n/2 \rceil \geq n/2$, we obtain
$$
  \mathbb{P}_x(\tau_T > n) \leq
  C_A(x)\, \beta_A^{-1/2}\, (\beta_A^{1/2})^n +
  C_T\, (\beta_T^{1/2})^n.
$$
Finally, setting $\beta = \max(\beta_A, \beta_T)^{1/2}$ and
$C(x) = C_A(x)\, \beta_A^{-1/2} + C_T$, we conclude that
$\mathbb{P}_x(\tau_T > n) \leq C(x)\, \beta^n$.
\end{proof}

When $A$ is finite and its prefactors are controlled in expectation along
runs, this lemma yields the following theorem: for every
$T \subseteq S$, the set $T \cup \widetilde{T}$ is geometrically
reachable. Unlike Theorem~6.1 in~\cite{AbdullaBenHendaMayrSandbergATVA2006},
whose eager attractor only controls runs starting in $A$, our bound holds
from every initial state, which, to the best of our knowledge, had not
been established before.

\begin{restatable}{theorem}{ThmFiniteAttractorPWSTS}
\label{thm:finite-attractor-pWSTS}
Let $(S, P)$ be a Markov chain, $A \subseteq S$ non-empty and finite, and
$T \subseteq S$. Assume that $A$ is geometrically reachable, with
$C_A \colon S \to \mathbb{R}_{>0}$ and $\beta_A \in (0,1)$, and that
there exists
$f \colon \mathbb{R}_{>0} \times \mathbb{N} \to \mathbb{R}_{>0}$
such that $\mathbb{E}_x[C_A(X_k)] \leq f(C_A(x), k)$ for every $x \in S$
and $k \in \mathbb{N}$. Then $T \cup \widetilde{T}$ is geometrically
reachable.
\end{restatable}
To prove this theorem, we first establish the following auxiliary lemma.

\begin{lemma}
\label{lem:mgf-exp-tail}
Let $Y$ be an $\Nat \cup \{\infty\}$-valued random variable with $C > 0$ and
$\beta \in (0,1)$ such that for all $n \in \Nat$,
$\Prob(Y > n) \leq C \beta^n$. Then for every
$s \in (0, \log(1/\beta))$, $\Exp[e^{sY}]
  \leq 1 + C\,\frac{e^s - 1}{1 - \beta e^s}$.

\end{lemma}

\begin{proof} Let $Y$ be an $\Nat \cup \{\infty\}$-valued random variable with $C > 0$ and
$\beta \in (0,1)$ such that, for all $n \in \Nat$,
$\Prob(Y > n) \leq C \beta^n$. Since $\Prob(Y=\infty)=\lim_nP(Y>n)\le \lim_nC\beta^n=0$, the variable $Y$ is almost surely finite. Then, for an almost surely finite $\Nat$-valued $Y$, the telescoping identity
$e^{sY} = 1 + (e^s - 1)\sum_{n \geq 0} e^{sn}\mathbf{1}_{Y > n}$ holds. Taking expectations and applying Tonelli,
\[
  \Exp[e^{sY}]
  = 1 + (e^s-1)\sum_{n \geq 0} e^{sn}\Prob(Y > n)
  \leq 1 + C(e^s-1)\sum_{n \geq 0}(\beta e^s)^n
  = 1 + C\frac{e^s-1}{1-\beta e^s},
\]
the series converging since $\beta e^s < 1$.
\end{proof}

Now, we prove the theorem. To this end, we first prove that $T\cup\widetilde T$ is uniformly geometrically reachable with respect to $A$, and then invoke the above lemma to obtain global geometric reachability.
\begin{proof}
If $T$ is empty, the claim is trivially true. Assume now that $T \neq \emptyset$. 

By Lemma~\ref{thm:eagerness-transitivity}, it suffices to show that $T \cup \widetilde{T}$ is a uniformly geometrically reachable set with respect to $A$. Put $A_0 = A \setminus (T \cup \widetilde{T})$. For $a \in A \cap (T \cup \widetilde{T})$, we have $\tau_{T \cup \widetilde{T}} = 0$ $\mathbb{P}_a$-a.s., so if $A_0 = \emptyset$ the claim is immediate; assume therefore $A_0 \neq \emptyset$. 
For each $a \in A_0 \subseteq \Pred^*(T)$, there is a finite run from $a$ to $T$ of length $k_a \in \Nat$; and since $a \notin T$, $k_a \geq 1$. As $A_0$ is finite, we set $\ell_0 = \max_{a \in A_0} k_a \geq 1$. Moreover, for each $a \in A_0$, we have: $\mathbb{P}_a(\tau_{T \cup \widetilde T} \leq \ell_0) \geq \mathbb{P}_a(\tau_T \leq \ell_0) \geq \prod_{1\leq j \leq k_a} P(y_j^a) (y_{j+1}^a) > 0$, so we can define a minimal success probability over $A_0$: $\delta = \min_{a \in A_0} \mathbb{P}_a(\tau_{T \cup \widetilde T} \leq \ell_0) > 0$. Set $\delta_0 = \min(\delta,1/2)$ and fix $a \in A$.

We define a sequence of return times by $\sigma_0 = 0$ and, for all $k \in \mathbb{N}$, $\sigma_{k+1} = \inf\{n \geq \sigma_k + \ell_0 : X_n \in A\}$. The hypothesis (1)  ensures that from every $y$, $\tau_A < \infty$ $\mathbb{P}_y$-a.s.\, meaning the times $\sigma_k$ are a.s.\ finite and strictly increasing (since $\ell_0 \geq 1$). Moreover, it satisfies $X_{\sigma_k} \in A$. Let $K = \inf\{k \in \mathbb{N} : \tau_{T \cup \widetilde{T}} \leq \sigma_k + \ell_0\}$ be the number of attempts. We prove by induction on $k \in \mathbb{N}$ that $\mathbb{P}_a(K > k) \leq (1-\delta_0)^k$.

For $k=0$, we have $\mathbb{P}_a(K > 0) \leq 1 = (1-\delta_0)^0$. 

For the induction step, fix $k \in \mathbb{N}_+$ and assume the
bound holds for $k-1$. By definition of $K$, we have the equality of events
$\{K > k\} = \{K > k-1\} \cap \{\tau_{T \cup \widetilde{T}} > \sigma_k +
\ell_0\}$. On the event $\{\tau_{T \cup \widetilde T} > \sigma_k\}$ we have
the identity $\tau_{T \cup \widetilde T} = \sigma_k + \tau_{T \cup \widetilde
T} \circ \theta_{\sigma_k}$. Consequently we only
have the pointwise inequality
$$
  \mathbf{1}_{\{\tau_{T \cup \widetilde T} > \sigma_k + \ell_0\}}
  \leq \mathbf{1}_{\{\tau_{T \cup \widetilde T} \circ \theta_{\sigma_k} >
  \ell_0\}},
$$
which suffices for our purposes. Since the event $\{K > k-1\}$ depends only
on the trajectory up to time $\sigma_{k-1} + \ell_0 \leq \sigma_k$, its
indicator function is $\mathcal{F}_{\sigma_k}$-measurable. By expressing the
probability as an expectation, using the above inequality, and conditioning
on $\mathcal{F}_{\sigma_k}$, we apply the strong Markov property at the
stopping time $\sigma_k$:
$$ \mathbb{P}_a(K > k) \leq \mathbb{E}_a \left[ \mathbf{1}_{\{K > k-1\}}
\mathbf{1}_{\{\tau_{T \cup \widetilde{T}} \circ \theta_{\sigma_k} >
\ell_0\}} \right] $$
$$ = \mathbb{E}_a \left[ \mathbb{E}_a \left[ \mathbf{1}_{\{K > k-1\}}
\mathbf{1}_{\{\tau_{T \cup \widetilde{T}} \circ \theta_{\sigma_k} > \ell_0\}}
\mid \mathcal{F}_{\sigma_k} \right] \right] $$
$$ = \mathbb{E}_a \left[ \mathbf{1}_{\{K > k-1\}}
\mathbb{E}_{X_{\sigma_k}} \left[ \mathbf{1}_{\{\tau_{T \cup \widetilde{T}} >
\ell_0\}} \right] \right] $$
$$ = \mathbb{E}_a \left[ \mathbf{1}_{\{K > k-1\}} \mathbb{P}_{X_{\sigma_k}}
(\tau_{T \cup \widetilde{T}} > \ell_0) \right] $$

By construction $X_{\sigma_k} \in A$ (regardless of whether $K > k-1$ holds),
so we may bound the inner probability uniformly over $A$ rather than over
$A_0$: if $X_{\sigma_k} \in T \cup \widetilde T$ then
$\mathbb{P}_{X_{\sigma_k}}(\tau_{T \cup \widetilde T} > \ell_0) = 0 \leq
1-\delta_0$ trivially, and if $X_{\sigma_k} \in A_0$ then the minimal
success-probability estimate gives $\mathbb{P}_{X_{\sigma_k}}(\tau_{T \cup
\widetilde{T}} > \ell_0) \leq 1 - \delta_0$ as before. In both cases,
$$
  \max_{a' \in A} \mathbb{P}_{a'}(\tau_{T \cup \widetilde{T}} > \ell_0)
  \leq 1 - \delta_0.
$$
Substituting this uniform bound back into the expectation gives:
$$ \mathbb{P}_a(K > k) \leq \mathbb{E}_a \left[ \mathbf{1}_{\{K > k-1\}}
(1 - \delta_0) \right] = (1 - \delta_0) \mathbb{P}_a(K > k-1) $$

Using the induction hypothesis, we obtain $\mathbb{P}_a(K > k) \leq (1-\delta_0)(1-\delta_0)^{k-1} = (1-\delta_0)^k$, which concludes the induction and ensures that $\mathbb{P}_a(K = \infty)=0$, thus $\tau_{T \cup \widetilde{T}} \leq \sigma_K + \ell_0 $ $\mathbb{P}_a$-a.s. 

Now, fix $s \in (0, \log(1/\beta_A))$ to bound the exponential moments of $\sigma_k$.  By definition, for all $k \in \mathbb{N}_+$, $\sigma_k = \sigma_{k-1} + \sigma_1 \circ \theta_{\sigma_{k-1}}$ and $\sigma_1 = \ell_0 + \tau_A \circ \theta_{\ell_0}$. Applying the strong Markov property at the stopping time $\sigma_{k-1}$ gives, for all $k \in \mathbb{N}_+$, 
$$ \mathbb{E}_a[e^{s\sigma_k}] = \mathbb{E}_a \left[ \mathbb{E}_a \left[ e^{s\sigma_{k-1}} e^{s(\sigma_1 \circ \theta_{\sigma_{k-1}})} \mid \mathcal{F}_{\sigma_{k-1}} \right] \right] = \mathbb{E}_a \left[ e^{s\sigma_{k-1}} \mathbb{E}_{X_{\sigma_{k-1}}}[ e^{s\sigma_1} ] \right] $$
For any starting state $a' \in A$, applying the Markov property at the deterministic time $\ell_0$ and using the moment generating function bound from Lemma~\ref{lem:mgf-exp-tail} yields 
$$ \mathbb{E}_{a'}[e^{s\sigma_1}] = e^{s\ell_0} \mathbb{E}_{a'} \left[ \mathbb{E}_{X_{\ell_0}}[e^{s\tau_A}] \right] \leq e^{s\ell_0} \left( 1 +  \frac{e^s - 1}{1 - \beta_A e^s} \mathbb{E}_{a'}[C_A(X_{\ell_0})] \right) $$
By assumption (2) of the theorem,  $\mathbb{E}_{a'}[C_A(X_{\ell_0})] \leq f(C_A(a'), \ell_0)$. Since $A$ is a finite set, we can take the supremum over all $a' \in A$ and define a valid, finite uniform bound 
$$ M(s) = e^{s\ell_0} \left( 1 + \frac{e^s - 1}{1 - \beta_A e^s} \max_{a' \in A} (f(C_A(a'), \ell_0) )\right) > 1 $$
Because $X_{\sigma_{k-1}} \in A$, we have $\mathbb{E}_{X_{\sigma_{k-1}}}[ e^{s\sigma_1} ] \leq M(s)$, which directly yields the recursive inequality $\mathbb{E}_a[e^{s\sigma_k}] \leq M(s) \mathbb{E}_a[e^{s\sigma_{k-1}}]$. Iterating this recurrence relation down to the base case $k=0$, where $\sigma_0 = 0$ and thus $\mathbb{E}_a[e^{s\sigma_0}] = 1$, we obtain the product bound $\mathbb{E}_a[e^{s\sigma_k}] \leq M(s)^k$. By Markov's inequality, this immediately implies for all $n\in\Nat$, $\mathbb{P}_a(\sigma_k > n) \leq e^{-sn} M(s)^k$. 

We now turn to the final step. Fix $n \in \mathbb{N}$. We first justify, for every $k \in \mathbb{N}$, the inclusion $\{\tau_{T \cup \widetilde T} > n\} \subseteq \{K > k\} \cup \{\sigma_k > n - \ell_0\}$. Let $k \in \Nat$, since $(\sigma_j)_j$ is strictly increasing, $K \leq k$ implies $\sigma_K \leq \sigma_k$, hence $\tau_{T \cup \widetilde T} \leq \sigma_K + \ell_0 \leq \sigma_k + \ell_0$ by definition of $K$; thus, $\{\tau_{T \cup \widetilde T} > \sigma_k + \ell_0\} \subseteq \{K > k\}$. Now suppose $\tau_{T \cup \widetilde T} > n$. If $\sigma_k \leq n - \ell_0$, then $\tau_{T \cup \widetilde T} > n \geq \sigma_k + \ell_0$, so the previous inclusion gives $K > k$. Otherwise $\sigma_k > n - \ell_0$ directly. In either case one of the two events on the right occurs, which proves the inclusion. Then, applying, for all $k \in \Nat$, $\Prob_a(K > k) \leq (1-\delta_0)^k$ and the exponential Markov inequality $\Prob_a(\sigma_k > n - \ell_0) \leq e^{-s(n - \ell_0)}\Exp_a[e^{s\sigma_k}] \leq e^{-s(n - \ell_0)} M(s)^k$, the union bound yields,
$$
  \Prob_a(\tau_{T \cup \widetilde T} > n)
  \leq (1-\delta_0)^k + e^{-s(n-\ell_0)} M(s)^k.
$$

The inequality holds for every $k$, so we may specialize $k$ to a value depending on $n$. The first term decreases in $k$ while the second increases (as $M(s) > 1$), so we balance them by setting $k = \lfloor \theta n \rfloor$ with $\theta = s/(2\log M(s))$, which is well defined since $M(s) > e^{s\ell_0} > 1$. For the first term, $\lfloor \theta n \rfloor \geq \theta n - 1$ and $\delta_0 \leq 1/2$ (so $1-\delta_0 \geq 1/2$) give
$$
  (1-\delta_0)^{\lfloor \theta n \rfloor}
  \leq (1-\delta_0)^{\theta n - 1}
  = (1-\delta_0)^{-1}\big[(1-\delta_0)^{\theta}\big]^n
  \leq 2\,\big[(1-\delta_0)^{\theta}\big]^n.
$$
For the second term, $\lfloor \theta n \rfloor \leq \theta n$ and $M(s) > 1$ give $M(s)^{\lfloor \theta n \rfloor} \leq M(s)^{\theta n}$, and by the choice of $\theta$,
$$
  M(s)^{\theta}
  = \exp\!\big(\theta \log M(s)\big)
  = \exp\!\left(\frac{s}{2\log M(s)}\log M(s)\right)
  = e^{s/2},
$$
so $e^{-s(n-\ell_0)} M(s)^{\theta n} = e^{s\ell_0}\,(e^{-s/2})^n$. Substituting both estimates at $k = \lfloor \theta n \rfloor$ gives
$$
  \Prob_a(\tau_{T \cup \widetilde T} > n)
  \leq 2\big[(1-\delta_0)^{\theta}\big]^n + e^{s\ell_0}\big(e^{-s/2}\big)^n.
$$
Setting $\beta' = \max\!\big((1-\delta_0)^{\theta},\, e^{-s/2}\big) \in (0,1)$ bounds each term by its prefactor times $(\beta')^n$, and summing the prefactors yields
$$
  \Prob_a(\tau_{T \cup \widetilde T} > n) \leq C'(\beta')^n,
  \qquad C' = 2 + e^{s\ell_0},
$$
uniformly over $a \in A$. The choice $s = \tfrac12\log(1/\beta_A)$ ensures $s \in (0, \log(1/\beta_A))$, as required for the moment bound defining $M(s)$ (since $\beta_A e^s = \beta_A^{1/2} < 1$), and gives the closed form $e^{-s/2} = \beta_A^{1/4}$. This establishes that $T \cup \widetilde{T}$ is uniformly geometrically
reachable with respect to $A$, with $C_T = C'$ and $\beta_T = \beta'$,
and Lemma~\ref{thm:eagerness-transitivity} concludes, $A$ being
geometrically reachable by assumption.
\end{proof}

\begin{remark}
The assumption on $f$ in Theorem~\ref{thm:finite-attractor-pWSTS} holds
in particular when $C_A$ grows in a controlled way along runs: if
$C_A(y) \leq f(C_A(x), k)$ whenever $\mathbb{P}_x(X_k = y) > 0$, taking
expectations yields it. This is the case for pLCS, where each
transition appends a bounded number of messages. The formulation in
expectation is, however, much more permissive: the prefactors may
explode along some runs, as long as these runs are improbable enough to
compensate. Probabilities thus absorb what no worst-case bound could,
and this is essential for infinitely branching models such as the
$\nu$-pLCS (see below).
\end{remark}

\begin{restatable}{proposition}{PropCalculus}
\label{prop:calculus}
Let $(\mathcal{S}, P)_{\mathrm{eff}}$ be an effective pWSTS and
$A, B \subseteq S$ non-empty and finite. Assume the hypotheses of Theorem~\ref{thm:finite-attractor-pWSTS} with
$T = {\uparrow}B$, where $C_A \colon S \to \mathbb{Q}_{>0}$ and
$f \colon \mathbb{Q}_{>0} \times \mathbb{N} \to \mathbb{Q}_{>0}$ are
computable, and $\beta_A \in (0,1) \cap \mathbb{Q}$ is given. Then one can compute
$\bar\beta \in \mathbb{Q} \cap (0,1)$ and
$\bar C \colon S \to \mathbb{Q}_{\geq 1}$ for the geometric reachability
of $T \cup \widetilde{T}$.
\end{restatable}

\begin{proof}
We retain the notation of the proof of
Theorem~\ref{thm:finite-attractor-pWSTS}, which remains valid for any
$\ell_0 \geq 1$ such that every $a \in A_0$ reaches $T$ within $\ell_0$
steps, and any rational $\delta_0$ with $0 < \delta_0 \leq \min(\delta, 1/2)$.
The backward coverability algorithm returns a finite basis of
$\Pred^*({\uparrow}B)$ and an index $\ell$ such that
$\Pred^*({\uparrow}B) = \Pred^{\leq \ell}({\uparrow}B)$. This basis allows us
to compute $A_0$, and we take $\ell_0 = \max(\ell, 1)$. If
$A_0 = \emptyset$, we set $\delta_0 = 1/2$. Otherwise, for each
$a \in A_0$, running $\mathcal{A}_2$ at horizon $\ell_0$ with decreasing
precision until the returned value is positive yields a rational
$p_a > 0$ with $p_a \leq \mathbb{P}_a(\tau_{{\uparrow}B} \leq \ell_0)$; this
terminates since this probability is positive. We set
$\delta_0 = \min(1/2, \min_{a \in A_0} p_a)$.

The constants $s$, $M(s)$, $\theta$, $C'$, $\beta'$, and then $\beta$ and
$C(x)$, are obtained from $\beta_A$, $\ell_0$, $\delta_0$, $C_A(x)$ and
$\max_{a \in A} f(C_A(a), \ell_0)$ using $+$, $\times$, $/$, $\exp$, $\log$
and $\max$; they are therefore computable reals. Since the bound
$\mathbb{P}_x(\tau_{T \cup \widetilde T} > n) \leq C(x)\beta^n$ remains
valid when $C(x)$ and $\beta$ are replaced by larger values, we take for
$\bar C(x)$ a rational upper approximation of $C(x)$, and for $\bar\beta$ a
rational upper approximation of $\beta$ with precision fine enough to
ensure $\bar\beta < 1$, which exists since $\beta < 1$.
\end{proof}

\subsection{Building a finite geometrically reachable set}

To apply the preceding theorems, one must first identify a geometrically reachable set. This subsection is dedicated to providing constructive methods to build such a set and to compute its parameters. The tools used in this section are drawn from the drift theory of
Markov chains~\cite{Hajek1982,meyn1993markov}.

\begin{restatable}{lemma}{LemHajekDrift}
\label{lem:hajek-drift}
Let $(S, P)$ be a Markov chain, $w \colon S \to \mathbb{Q}_{\geq 1}$ a
weight function, $A \subseteq S$ non-empty and $\beta \in (0,1)$ such
that $\mathbb{E}_x[w(X_1)] \leq \beta\, w(x)$ for every $x \notin A$.
Then $A$ is geometrically reachable, with $C_A = w$ and 
$\beta_A = \beta$.
\end{restatable}
\begin{proof}
If $x \in A$, then $\tau_A = 0$ $\mathbb{P}_x$-a.s., so
$\mathbb{P}_x(\tau_A > n) = 0 \leq w(x)\beta^n$ and the claim is trivial.
Assume now $x \notin A$. Consider the process
$$
  M_n = w(X_{\min(n, \tau_A)})\,\beta^{-\min(n, \tau_A)},
  \qquad n \geq 0.
$$
For each $n \geq 0$, $M_n > 0$ and $M_n$ is $\mathcal{F}_n$-measurable. We
claim that $(M_n)_{n \geq 0}$ is a supermartingale, i.e.\
$\mathbb{E}_x[M_{n+1} \mid \mathcal{F}_n] \leq M_n$ for all $n \geq 0$.

The events $\{\tau_A \leq n\}$ and $\{\tau_A > n\}$ are
$\mathcal{F}_n$-measurable, so their indicators factor out of
$\mathbb{E}_x[\,\cdot \mid \mathcal{F}_n]$, and we split
$$
  \mathbb{E}_x[M_{n+1} \mid \mathcal{F}_n]
  = \mathbf{1}_{\{\tau_A \leq n\}}\,\mathbb{E}_x[M_{n+1} \mid \mathcal{F}_n]
  + \mathbf{1}_{\{\tau_A > n\}}\,\mathbb{E}_x[M_{n+1} \mid \mathcal{F}_n].
$$
On $\{\tau_A \leq n\}$, the process is stopped: $\min(n+1, \tau_A) =
\min(n, \tau_A) = \tau_A$, so $M_{n+1} = w(X_{\tau_A})\beta^{-\tau_A} = M_n$,
which is $\mathcal{F}_n$-measurable. Hence
$$
  \mathbf{1}_{\{\tau_A \leq n\}}\,\mathbb{E}_x[M_{n+1} \mid \mathcal{F}_n]
  = \mathbf{1}_{\{\tau_A \leq n\}}\,M_n.
$$
On $\{\tau_A > n\}$, we have $X_n \notin A$ and $\min(n+1, \tau_A) = n+1$,
so $M_{n+1} = w(X_{n+1})\beta^{-(n+1)}$ and
$$
  \mathbf{1}_{\{\tau_A > n\}}\,\mathbb{E}_x[M_{n+1} \mid \mathcal{F}_n]
  = \mathbf{1}_{\{\tau_A > n\}}\,\beta^{-(n+1)}\,
    \mathbb{E}_x[w(X_{n+1}) \mid \mathcal{F}_n].
$$
To evaluate the conditional expectation, observe that $X_{n+1} = X_1 \circ
\theta_n$, so that $w(X_{n+1}) = \Phi \circ \theta_n$ with $\Phi := w \circ
X_1$, a non-negative measurable functional on $\Omega$. The Markov property applied at the deterministic time $n$ then
gives
$$
  \mathbb{E}_x[w(X_{n+1}) \mid \mathcal{F}_n]
  = \mathbb{E}_x[\Phi \circ \theta_n \mid \mathcal{F}_n]
  = \mathbb{E}_{X_n}[\Phi]
  = \mathbb{E}_{X_n}[w(X_1)].
$$
Since $X_n \notin A$ on $\{\tau_A > n\}$, the drift hypothesis applies at
the state $X_n$ and yields $\mathbb{E}_{X_n}[w(X_1)] \leq \beta\, w(X_n)$,
whence
$$
  \mathbf{1}_{\{\tau_A > n\}}\,\mathbb{E}_x[M_{n+1} \mid \mathcal{F}_n]
  \leq \mathbf{1}_{\{\tau_A > n\}}\,\beta^{-(n+1)}\,\beta\, w(X_n)
  = \mathbf{1}_{\{\tau_A > n\}}\,M_n.
$$
Summing the two contributions,
$$
  \mathbb{E}_x[M_{n+1} \mid \mathcal{F}_n]
  \leq \mathbf{1}_{\{\tau_A \leq n\}}\,M_n
     + \mathbf{1}_{\{\tau_A > n\}}\,M_n
  = M_n,
$$
so $(M_n)$ is a supermartingale.

Taking expectations and iterating, $\mathbb{E}_x[M_n] \leq \mathbb{E}_x[M_0]
= w(x)$ for all $n \geq 0$. On the event $\{\tau_A > n\}$ we have
$\min(n, \tau_A) = n$, hence $M_n = w(X_n)\beta^{-n} \geq \beta^{-n}$ since
$w \geq 1$; thus $\{\tau_A > n\} \subseteq \{M_n \geq \beta^{-n}\}$.
Markov's inequality then gives
$$
  \mathbb{P}_x(\tau_A > n)
  \leq \mathbb{P}_x\big(M_n \geq \beta^{-n}\big)
  \leq \frac{\mathbb{E}_x[M_n]}{\beta^{-n}}
  \leq w(x)\,\beta^n,
$$
which is the claimed bound.
\end{proof}

\begin{restatable}{proposition}{PropPWSTSAffineDrift}
\label{thm:pWSTS-affine-drift}
Let $(S, P)$ be a Markov chain and $w \colon S \to  \mathbb{Q}_{\geq 1}$ be a weight function. Assume there exist constants $\lambda \in (0,1)$ and $b \geq 1$ such that for all $x \in S$,
$ \mathbb{E}_x[w(X_1)] \leq \lambda\, w(x) + b. $ Then, if $A = \left\{x \in S : w(x) \leq \frac{2b}{1-\lambda}\right\}$ is not empty, $A$ is a geometrically reachable set with $C_A = w $ and $\beta_A=\frac{1+\lambda}{2}$.
\end{restatable}

\begin{proof}
Let $R = \frac{2b}{1-\lambda}$ and $\beta_A = \frac{1+\lambda}{2}$. Since $\lambda \in (0,1)$, it follows that $\beta_A \in (0,1)$. 

For any state $x \notin A$, we have by definition $w(x) > R$. This implies that 
$ \frac{b}{w(x)} < \frac{b}{R} = \frac{1-\lambda}{2}. $

Using the affine drift hypothesis for $x \notin A$, we obtain:
$$ \frac{\mathbb{E}_x[w(X_1)]}{w(x)} \leq \lambda + \frac{b}{w(x)} < \lambda + \frac{1-\lambda}{2} = \frac{1+\lambda}{2} = \beta_A. $$

Hence, $\mathbb{E}_x[w(X_1)] \leq \beta_A\, w(x)$ for all $x \notin A$. Thus the multiplicative drift condition of Lemma~\ref{lem:hajek-drift} holds on $S\setminus A$.
$$ \mathbb{P}_x(\tau_A > n) \leq w(x)\, \beta_A^n = w(x)\,\left(\frac{1+\lambda}{2}\right)^n $$
for all $x \in S$ and $n \geq 0$, which concludes the proof.
\end{proof}

\begin{remark}
Note that for any $b' \ge b$, the corresponding target set $A'$ contains $A$. Therefore, the assumption that $A$ is non-empty is satisfied in practice provided the rest is satisfied: one can simply pick an arbitrary state $x_0 \in S$ and take $b' = \max(b, w(x_0)) $. Since $\lambda \in (0,1)$, we have $\frac{2b'}{1-\lambda} > w(x_0)$, which guarantees that $x_0 \in A$.\end{remark}

Depending on the definition of the weight function $w$, the set $A$ might not necessarily be finite. In practice, $w$ is designed precisely so that its sublevel sets, typically of the form $\{s \in S \mid w(s) \le K\}$ for a given constant $K$, are finite. For instance, in the case of pLCS (which we will discuss later), a natural weight function is the length of the configuration's word which leads to finite sublevel sets. Therefore, to maintain generality in our theoretical framework, we must explicitly state the finiteness of $A$ as an assumption.

\begin{restatable}{theorem}{ThmDriftHorizon}
\label{thm:drift-horizon}
Let $(S, P)$ be a Markov chain, $T \subseteq S$ and
$w \colon S \to \mathbb{Q}_{\geq 1}$ a weight function. Assume that there
exist $\lambda \in (0,1)$ and $b \geq 1$ such that
$\mathbb{E}_x[w(X_1)] \leq \lambda\, w(x) + b$ for every $x \in S$, and
that $A = \{x \in S : w(x) \leq \frac{2b}{1-\lambda}\}$ is finite and
non-empty. Then there exists
$\mathcal{N}_T \colon (\mathbb{Q} \cap (0,1)) \times S \to \mathbb{N}$,
such that, for every $(\varepsilon, x) \in (\mathbb{Q} \cap (0,1)) \times S$:
\begin{itemize}
  \item once $w(x)$ is known, the evaluation of
  $\mathcal{N}_T(\varepsilon, x)$ requires a constant (independent of $\varepsilon$ and $x$) number of elementary
  operations $(+, \times, /, \log)$,
  followed by a rational upper approximation and a ceiling;
  \item $\mathcal{N}_T(\varepsilon, x) \in n^*_T(\varepsilon, x)$;
  \item $\mathcal{N}_T(\varepsilon, x) = O\big(\log(1/\varepsilon) + \log w(x)\big)$
  as $\varepsilon \to 0$ and $w(x) \to +\infty$.
\end{itemize}
\end{restatable}

\begin{proof}
Let $R = \frac{2b}{1-\lambda}$. By Proposition~\ref{thm:pWSTS-affine-drift},
$A$ is geometrically reachable with $\beta_A = \frac{1+\lambda}{2}$
and $C_A=w$. Iterating the drift condition gives, for every
$x \in S$ and $k \in \mathbb{N}$,
$$
  \mathbb{E}_x[w(X_k)] \leq \lambda^k w(x) + b \sum_{i=0}^{k-1} \lambda^i
  \leq w(x) + \frac{b}{1-\lambda},
$$
by induction on $k$, using
$\mathbb{E}_x[w(X_k)] = \mathbb{E}_x[\mathbb{E}_{X_{k-1}}[w(X_1)]] \leq
\lambda\,\mathbb{E}_x[w(X_{k-1})] + b$. The hypotheses of
Theorem~\ref{thm:finite-attractor-pWSTS} thus hold with
$f(u, k) = u + \frac{b}{1-\lambda}$, which is non-decreasing in $u$. We
follow its proof and make the constants explicit.

Let
$A_0 = A \setminus (T \cup \widetilde{T})$. The proof of
Theorem~\ref{thm:finite-attractor-pWSTS} remains valid for any
$\ell_0 \geq 1$ such that every $a \in A_0$ reaches $T$ within $\ell_0$
steps with positive probability, and any $\delta_0 \in (0, 1/2]$ such that
$\delta_0 \leq \min_{a \in A_0} \mathbb{P}_a(\tau_{T \cup \widetilde T} \leq \ell_0)$.
We fix such $\ell_0$ and $\delta_0$. If $A_0 = \emptyset$, then
$\mathbb{P}_a(\tau_{T \cup \widetilde T} > n) = 0$ for every $a \in A$, and
the bounds below hold with $\ell_0 = 1$ and $\delta_0 = 1/2$.

Fix $s = \frac12 \log(1/\beta_A)$, so that
$e^s = \beta_A^{-1/2}$. The fractional term of the moment bound collapses:
$$
  \frac{e^s - 1}{1 - \beta_A e^s}
  = \frac{\beta_A^{-1/2} - 1}{1 - \beta_A^{1/2}} = \beta_A^{-1/2}.
$$
Since $w(a) \leq R$ for every $a \in A$ and $f$ is non-decreasing in its
first argument,
$\max_{a \in A} f(w(a), \ell_0) \leq f(R, \ell_0) = \frac{3b}{1-\lambda}$.
As the argument only uses an upper bound on the exponential moments, we
may take
$$
  M = \left(\frac{2}{1+\lambda}\right)^{\ell_0/2}
  \left(1 + \sqrt{\frac{2}{1+\lambda}}\;\frac{3b}{1-\lambda}\right) > 1,
$$
and $\theta = \frac{s}{2 \log M}$. The return-process constants are
$C' = 2 + \left(\frac{2}{1+\lambda}\right)^{\ell_0/2}$ and
$\beta' = \max\big((1-\delta_0)^{\theta}, \beta_A^{1/4}\big)$. Since
$\beta_A < \beta_A^{1/4} \leq \beta'$, the global rate of
Lemma~\ref{thm:eagerness-transitivity} simplifies to
$$
  \beta = \max(\beta_A, \beta')^{1/2}
  = \max\big((1-\delta_0)^{\theta/2}, \beta_A^{1/8}\big) \in (0,1),
$$
and the prefactors are
$$
  C(x) = w(x)\sqrt{\frac{2}{1+\lambda}}
  + \left(\frac{2}{1+\lambda}\right)^{\ell_0/2} + 2.
$$
Hence $\mathbb{P}_x(\tau_{T \cup \widetilde T} > n) \leq C(x)\,\beta^n$
for all $x \in S$ and $n \geq 0$.

Let
$v(\varepsilon, x) = \frac{\log(1/\varepsilon) + \log C(x)}{\log(1/\beta)}$,
which is non-negative since $C(x) \geq 1$ and $\varepsilon < 1$. Let
$q(\varepsilon, x) \in \mathbb{Q}$ be such that
$v(\varepsilon, x) \leq q(\varepsilon, x) \leq v(\varepsilon, x) + 1$, and
set $\mathcal{N}_T(\varepsilon, x) = \lceil q(\varepsilon, x) \rceil$.
For every $n \geq \mathcal{N}_T(\varepsilon, x)$, we have
$n \geq v(\varepsilon, x)$, hence $C(x)\,\beta^n \leq \varepsilon$, so
$\mathcal{N}_T(\varepsilon, x) \in n^*_T(\varepsilon, x)$. Once $w(x)$ is
known, $C(x)$ and $v(\varepsilon, x)$ are obtained by a fixed number of
operations among $(+, \times, /, \log)$, the constants depending only on
$\lambda$, $b$, $\ell_0$ and $\delta_0$. Finally,
$\mathcal{N}_T(\varepsilon, x) \leq v(\varepsilon, x) + 2$ and, since
$w \geq 1$, $C(x) \leq c\, w(x)$ with
$c = \sqrt{\frac{2}{1+\lambda}} + C'$, so
$\log C(x) \leq \log w(x) + \log c$, which gives
$\mathcal{N}_T(\varepsilon, x) = O\big(\log(1/\varepsilon) + \log w(x)\big)$.
\end{proof}

The explicit drift condition makes several quantities
of the proof of Theorem~\ref{thm:finite-attractor-pWSTS} simplify.
The remaining choices, such as the splitting of the horizon, are not
optimized and could be tuned to sharpen the constants. For effective pWSTS, all these constants can be
computed.

\begin{restatable}{proposition}{PropDriftHorizonComputable}
\label{prop:drift-horizon-computable}
Let $(\mathcal{S}, P)_{\mathrm{eff}}$ be an effective pWSTS and
$B \subseteq S$ finite. Assume the hypotheses of
Theorem~\ref{thm:drift-horizon} with $T = {\uparrow}B$, where $w$ is
computable, and
$\lambda \in (0,1) \cap \mathbb{Q}$, $b \in \mathbb{Q}_{\geq 1}$ and $A$
are given.
Then a function $\mathcal{N}_{\uparrow B}$ as in Theorem~\ref{thm:drift-horizon} is computable. 
\end{restatable}

\begin{proof}
It suffices to compute the constants $\ell_0$ and $\delta_0$ of the proof
of Theorem~\ref{thm:drift-horizon}. The backward coverability algorithm
returns a finite basis $B'$ of $\Pred^*({\uparrow}B)$ and an index $L$
such that $\Pred^*({\uparrow}B) = \Pred^{\leq L}({\uparrow}B)$. Membership
in ${\uparrow}B$ and ${\uparrow}B'$ being decidable, we compute
$A_0 = (A \cap {\uparrow}B') \setminus {\uparrow}B$, and every
$a \in A_0$ reaches ${\uparrow}B$ within $L$ steps with positive
probability, so we take $\ell_0 = \max(L, 1)$. For each $a \in A_0$,
running $\mathcal{A}_2$ at horizon $\ell_0$ with decreasing precision
until the returned value is positive yields a rational $p_a > 0$ with
$p_a \leq \mathbb{P}_a(\tau_{{\uparrow}B} \leq \ell_0)$ (this terminates
since this probability is positive). We set
$\delta_0 = \min(1/2, \min_{a \in A_0} p_a)$, or $\delta_0 = 1/2$ if
$A_0 = \emptyset$.

The constants $M$, $\theta$ and $\beta$ are then obtained from the
rationals $\lambda$, $b$, $\ell_0$, $\delta_0$ by finitely many operations
among $+$, $\times$, $/$, $\exp$, $\log$ and $\max$, and $C(x)$ from
$w(x)$ in the same way; they are computable reals, and so is
$v(\varepsilon, x)$. A rational $q(\varepsilon, x)$ with
$v(\varepsilon, x) \leq q(\varepsilon, x) \leq v(\varepsilon, x) + 1$ is
obtained by computing a rational $r$ with
$|r - v(\varepsilon, x)| \leq 1/2$ and setting $q(\varepsilon, x) = r + 1/2$.
\end{proof}

\begin{remark}
\label{rem:log-structural}
As noted in Remark~\ref{rem:min-log}, the logarithmic behaviour does not rely on computability: it stems from
the geometric decay itself. In particular, for a fixed initial state, dividing the precision by a
factor $c > 1$ increases the depth $\mathcal{N}_T$ by at most
$\log(c)/\log(1/\beta) + 2$. A much finer precision only requires a
small number of extra steps. Furthermore, the computability of
Proposition~\ref{prop:drift-horizon-computable} extends to any
$T \subseteq S$ for which membership in $T$ and in $\Pred^*(T)$ is
decidable, since, as noted in Section~\ref{sec:preli}, $\mathcal{A}_2$
then applies to $T$~\cite{fffy2026}.
\end{remark}

\subsection{$\nu$-pLCS}
\label{sec:applications}

We now apply our framework to pLCS and to $\nu$-pLCS, which generalize the
emission mechanism of pLCS: an emission may append several messages at once,
their number being drawn from a probability distribution $p(\nu)$
parameterized by $\nu$.
When $p(\nu)$ has unbounded support, the induced transition system is
infinitely branching, as for $\omega$-Petri nets~\cite{GeeraertsHPR15}.
Existing eager-attractor results for
pLCS~\cite{AbdullaBenHendaMayrSandbergATVA2006,BenHenda08} do not directly
cover this model: their GR-attractor argument requires that at most one
message be added per step. Using the total channel length plus one as
weight function, we show that any distribution $p(\nu)$ with finite
expectation yields a finite geometrically reachable set, which provides
$\mathcal{N}_{{\uparrow}B}$, for pLCS and $\nu$-pLCS alike.

\begin{definition}[$\nu$-pLCS]
\label{def:nu-pLCS}
A \emph{probabilistic lossy channel system with $\nu$-emissions}
($\nu$-pLCS) is a tuple $(Q, C, \Sigma, \Delta, W, \nu, p_\ell)$, where
$Q$, $C$ and $\Sigma$ are finite sets of control states, channels and
messages, $\nu \in \mathbb{Q}_{\ge0}$, $p_\ell \in (0,1) \cap \mathbb{Q}$ is the
loss probability, $\Delta$ is a finite set of rules of the forms
$q \xrightarrow{c\,!\,\nu} q'$, $q \xrightarrow{c\,?\,m} q'$ and
$q \xrightarrow{\tau} q'$, with $q, q' \in Q$, $c \in C$ and
$m \in \Sigma$, and $W \colon \Delta \to \mathbb{N}^+$ assigns a positive
weight to each rule.
\end{definition}

A configuration is a pair $x = (q, (w_c)_{c \in C}) \in Q \times (\Sigma^*)^C$,
and $|x| = \sum_{c \in C} |w_c|$ denotes its total number of messages. As
for pLCS (Section~\ref{sec:preli}), the system follows a \textit{loss-action
semantics}. Let $p(\nu)$ be a probability distribution on $\mathbb{N}$ with
finite expectation. A $\nu$-emission $q \xrightarrow{c\,!\,\nu} q'$ is
always enabled: it draws $M \sim p(\nu)$, then a word $w$ uniformly in
$\Sigma^M$, appends $w$ to channel $c$ and moves to $q'$. A reception
$q \xrightarrow{c\,?\,m} q'$ is enabled iff channel $c$ starts with $m$;
it removes this message and moves to $q'$. An internal action
$q \xrightarrow{\tau} q'$ is always enabled and moves to $q'$ without
modifying the channels.

Ordered by the componentwise subword relation on channel contents (with
equal control states) and operating under a \emph{loss-action semantics}, the transition system is strongly monotone, thereby forming a WSTS. We assume that every control state has at least one outgoing emission or
internal rule. Since these rules are always enabled, every configuration
obtained after the loss step has an enabled rule, and $\nu$-pLCS are
deadlock-free. As for pLCS, combining the loss probability $p_\ell$, the rule weights $W$,
the distribution $p(\nu)$ and the uniform law on $\Sigma^M$ yields a
compatible Markov kernel on configurations.

Indeed, for any pair of configurations $x = (q_x, w_x)$ and $y = (q_y, w_y)$, the transition probability $P(x)(y)$ is obtained by summing over all intermediate subwords $w' \sqsubseteq w_x$ that survive the initial loss step:
$$
  P(x)(y) = \sum_{w' \sqsubseteq w_x} \mathbb{P}_{\text{loss}}(w_x \to w') \cdot \mathbb{P}_{\text{rule}}((q_x, w') \to y) .
$$

Note that the loss probability $\mathbb{P}_{\text{loss}}(w_x \to w')$ is computable from $p_\ell \in (0,1) \cap \mathbb{Q}$~\cite{Schnoebelen2004}. Furthermore, the rule execution probability $\mathbb{P}_{\text{rule}}$ depends on the relative weights of the currently enabled rules. Let $\Delta(q_x, w')$ denote the set of enabled rules from the intermediate configuration, and $W_{\text{total}} = \sum_{\delta' \in \Delta(q_x, w')} W(\delta')$ be their cumulative weight. The transition probability then expands to:
$$
  \mathbb{P}_{\text{rule}}((q_x, w') \to y) = \sum_{\delta \in \Delta(q_x, w')} \frac{W(\delta)}{W_{\text{total}}} \cdot \mathbb{P}_\delta((q_x, w') \to y) .
$$

This sum is effectively computable as long as the burst length probability $\mathbb{P}(M=k) = p_k(\nu)$ is computable for any $k \geq 0$. For internal actions and receptions, the conditional probability $\mathbb{P}_\delta(q_x, w' \to y)$ evaluates trivially to $1$ or $0$. For a $\nu$-emission rule appending a word $u \in \Sigma^*$ to the channel, this probability is $p_{|u|}(\nu) \cdot |\Sigma|^{-|u|}$. Once $M$ is fixed, the payload may be generated in various ways: the drift results only depend on $\mathbb{E}[M]$, and the effectiveness proof only
requires that every word of an admissible length be emitted with positive probability.

\begin{remark}
\label{rem:nu-pLCS-semantics}
The drift results below only depend on the expected number of messages
appended by an emission. They therefore hold for any emission semantics,
whatever the way the payload is generated, as long as the number of
appended messages has finite expectation.
\end{remark}

For $p(\nu) = \delta_\nu$, the Dirac distribution at $\nu \in \mathbb{N}$,
each $\nu$-emission appends $\nu$ messages. In particular,
$p(\nu) = \delta_1$ yields single-message emissions as in pLCS, the emitted
message being drawn uniformly from $\Sigma$ instead of being fixed by the
rule; for $\nu \geq 2$, emissions are larger but still bounded by $\nu$.
With the Poisson distribution $p(\nu) = \mathcal{P}(\nu)$, $\nu > 0$, or the
geometric distribution $p(\nu) = \mathcal{G}(\nu)$ on $\mathbb{N}$,
$\nu \in (0,1)$, the emission length is finite but unbounded, with
expectation $\nu$ and $(1-\nu)/\nu$ respectively.

\begin{restatable}{theorem}{ThmNuPLCSEffectiveness}
\label{theoreme:nu-pLCS-effectiveness}
Let $\mathcal{L}$ be a $\nu$-pLCS equipped with its Markov kernel $P$.
Assume that $p(\nu)$ is computable, that membership in
$\mathrm{supp}(p(\nu))$ is decidable, and that one can decide, for every
$v \in \mathbb{N}$, whether $\mathrm{supp}(p(\nu))$ contains some
$v' \geq v$. Then $(\mathcal{L}, P)$ is an effective pWSTS.
\end{restatable}

\begin{proof}
As established previously, a $\nu$-pLCS is a pWSTS. We verify the effectiveness conditions.
\begin{itemize}
  \item \textbf{The relations $\sqsubseteq$ and $=$ are decidable}, since
  both amount to comparing control states for equality and channel contents
  for equality or for the subword relation.

  \item \textbf{$\to$ is decidable.} Since $p_\ell \in (0,1)$, every
  $w' \sqsubseteq w_x$ is reached by the loss step with positive
  probability. Hence $P(x)(y) > 0$ iff there exist $w' \sqsubseteq w_x$ and
  $\delta \in \Delta(q_x, w')$ such that
  $\mathbb{P}_\delta((q_x, w') \to y) > 0$. There are finitely many such
  pairs $(w', \delta)$, and each condition is decidable: for internal
  actions and receptions it is a syntactic check, and for a $\nu$-emission
  appending $u$ it amounts to $|u| \in \mathrm{supp}(p(\nu))$, which is
  decidable by assumption.

  \item \textbf{The Markov kernel $P$ is computable}: $P(x)(y)$ is a finite
  sum of products of $\mathbb{P}_{\text{loss}}(w_x \to w')$, computable from
  $p_\ell$, of rational weight ratios, and of the quantities
  $p_{|u|}(\nu) \cdot |\Sigma|^{-|u|}$, computable since $p(\nu)$ is.

  \item \textbf{Post effective enumerability.} Since $M \sim p(\nu)$ may
  have infinite support, $\Post(x)$ may be infinite. However,
  $S = Q \times (\Sigma^*)^C$ is countable: fixing orders on $Q$, $C$ and
  $\Sigma$, we define a computable bijection
  $\mathsf{gen} \colon \mathbb{N} \to S$ that lists the finite sets
  $Q \times \{w \in (\Sigma^*)^C \mid \sum_{c \in C} |w_c| = n\}$ for
  $n = 0, 1, 2, \dots$, each ordered lexicographically. Given $x$ and $m$,
  we compute $\mathrm{succ}(x,m)$ by scanning $y = \mathsf{gen}(0),
  \mathsf{gen}(1), \dots$, testing $x \to y$ (decidable), and returning the
  $m$-th configuration for which the test succeeds. This procedure
  terminates iff $m \leq |\Post(x)|$, hence $\Post$ is effectively
  enumerable.

  \item \textbf{Pred-basis effectivity.} Following
  \cite{geffroy:hal-02391841}, we define for each rule $\delta$ an operator
  $\mathrm{cpre}_\delta(x)$ computing a finite basis of the configurations
  reaching $\uparrow x$ via $\delta$. Since losses only remove messages, any
  configuration reaching $\uparrow x$ dominates one that reaches it without
  any loss; minimal predecessors can therefore be computed assuming no
  message is lost, and a basis of $\Pred(\uparrow x)$ is given by
  $\bigcup_{\delta \in \Delta} \mathrm{cpre}_\delta(x)$.
  For $x = (q_x, w_x)$:
  \begin{itemize}
    \item If $\delta = q \xrightarrow{\tau} q_x$: $\mathrm{cpre}_\delta(x)
    = \{(q, w_x)\}$.
    \item If $\delta = q \xrightarrow{c?m} q_x$: $\mathrm{cpre}_\delta(x) =
    \{(q, w')\}$ where $w'_c = m \cdot w_{x,c}$ and $w'_d = w_{x,d}$ for
    $d \neq c$.
    \item If $\delta = q \xrightarrow{c!\nu} q_x$: let $\ell = |w_{x,c}|$.
    Since every word of length $v \in \mathrm{supp}(p(\nu))$ is emitted with
    positive probability, an emission of length $v$ covers the suffix of
    $w_{x,c}$ of length $\min(v, \ell)$ 
    By assumption, we can decide whether
    some $v' \in \mathrm{supp}(p(\nu))$ satisfies $v' \geq \ell$.
    \begin{itemize}
      \item If so, the emission can cover the whole target word, and we set
      $w'_c = \epsilon$.
      \item Otherwise, $\mathrm{supp}(p(\nu)) \subseteq \{0, \dots, \ell-1\}$,
      and $l = \max \mathrm{supp}(p(\nu))$ is computed by testing membership
      of $\ell-1, \ell-2, \dots, 0$ in turn (the support is non-empty). We
      set $w'_c$ to be the prefix of $w_{x,c}$ of length $\ell - l$.
    \end{itemize}
    In both cases, $\mathrm{cpre}_\delta(x) = \{(q, w')\}$ with $w'_c$ as
    above and $w'_d = w_{x,d}$ for $d \neq c$.
  \end{itemize}
  Each $\mathrm{cpre}_\delta(x)$ is computable, and since $\Delta$ is
  finite, their union is a computable finite basis of $\Pred(\uparrow x)$.
\end{itemize}
\end{proof}

These assumptions hold for usual distributions, such as the Poisson and
geometric distributions with rational parameter: their values are
computable, and their support is $\mathbb{N}$, which makes the last two
assumptions trivial.

\begin{remark}
Since for a computable real, positivity is
only semi-decidable, computability of $p(\nu)$ does not make membership in
$\mathrm{supp}(p(\nu))$ decidable.
\end{remark}
For $x \in Q \times (\Sigma^*)^C$, we define the weight function as $w(x) = |x| + 1$.

\begin{restatable}{lemma}{LemPLCSDrift}
\label{lem:pLCS-drift}
Let $\mathcal{L} = (Q, C, \Sigma, \Delta, W, \nu, p_\ell)$ be a $\nu$-pLCS equipped with its Markov kernel $P$. There exist $\lambda \in (0,1)\cap \mathbb{Q}$ and $b \in \mathbb{Q}_{\geq1}$ such that for all $x \in Q \times (\Sigma^*)^C$, $\mathbb{E}_x[w(X_1)] \leq \lambda\, w(x) + b$.
\end{restatable}

\begin{proof}
Let $M \sim p(\nu)$ be a random variable, and let $m = \mathbb{E}[M]$. Let $x \in Q \times (\Sigma^*)^C$. We have $|X_1| = N_{loss} + N_{rule}$, where $N_{loss}$ is a random variable representing the number of messages in the FIFO channels after the loss transition, and $N_{rule}$ is a random variable representing the number of messages added (or $-1$ if deleted and $0$ for internal transition) to the FIFO channels after the rule transition. Since $N_{loss} \sim \mathrm{Binomial}(|x|, 1-p_\ell)$, we have $\mathbb{E}_x[N_{loss}] = (1-p_\ell)|x|$. Furthermore, since $N_{rule} \leq M$, we obtain $\mathbb{E}_x[N_{rule}] \leq m$. By linearity of expectation, $\mathbb{E}_x[|X_1|] \leq (1-p_\ell)|x| + m$. Thus,
$$
\mathbb{E}_x[w(X_1)] \leq (1-p_\ell)|x| + m + 1 = (1-p_\ell)w(x) - (1 - p_\ell) + m + 1 \leq \lambda\, w(x) + b
$$
with $\lambda = 1 - p_\ell$ and any rational $b \ge \max(m + p_\ell,1)$.
\end{proof}

\begin{restatable}{corollary}{CorNuPLCSDecisive}
\label{cor:nu-pLCS-decisive}
Let $\mathcal{L}$ be a $\nu$-pLCS equipped with its Markov kernel $P$, and
$T \subseteq S$. Then $(\mathcal{L}, P)$ is decisive with respect to $T$.
\end{restatable}

\begin{proof}
The case $T = \emptyset$ is trivial. Otherwise, by Lemma~\ref{lem:pLCS-drift},
the proof of Theorem~\ref{thm:drift-horizon} applies and shows that
$T \cup \widetilde{T}$ is geometrically reachable.
Proposition~\ref{prop:eager-decisive} concludes.
\end{proof}

One can furthermore consider the set $A = \left\{x \in S : w(x) \leq \frac{2b}{1-\lambda}\right\}$, which is moreover finite and not empty.

\begin{restatable}{proposition}{PropPLCSEmissionHorizon}
\label{cor:pLCS-k-emission-horizon}
Let $(\mathcal{L},P)$ be a $\nu$-pLCS equipped with $P$ satisfying the assumptions of
Theorem~\ref{theoreme:nu-pLCS-effectiveness}, such that
$\mathbb{E}[M]$, for $M \sim p(\nu)$, is a computable real number, and let
$B \subseteq S$ be a finite set. Then a function $\mathcal{N}_{\uparrow B}$ as in Theorem~\ref{thm:drift-horizon} is computable. 
\end{restatable}

\begin{proof}
Since $\mathbb{E}[M]$ is a computable real, we compute once a rational
$\bar m \geq \mathbb{E}[M]$. By Lemma~\ref{lem:pLCS-drift}, the system
satisfies the affine drift condition with $\lambda = 1 - p_\ell \in (0,1)$
and $b = \max(\bar m + p_\ell, 1)$. Let $R = \frac{2b}{1-\lambda}$ and
$A = \{x \in S : w(x) \leq R\}$. Since $w(x) = |x| + 1$, $A$ consists of
the configurations with at most $\lfloor R \rfloor - 1$ messages in
total. It is non-empty, as $R \geq 2$ and configurations with empty
channels have weight $1$. It is also finite, since $Q$, $C$ and $\Sigma$
are finite, and its elements can be listed explicitly by enumerating,
for each control state and each total length
$n \leq \lfloor R \rfloor - 1$, all tuples of channel contents. The
assumptions of Proposition~\ref{prop:drift-horizon-computable} are
therefore satisfied, which yields a computable function
$\mathcal{N}_{{\uparrow}B}$. Finally, since $w(x) = |x| + 1$, the bound
$O\big(\log(1/\varepsilon) + \log w(x)\big)$ reads
$O\big(\log(1/\varepsilon) + \log(|x| + 1)\big)$.
\end{proof}

By Remark~\ref{rem:nu-pLCS-semantics}, the drift argument applies
verbatim to pLCS, with $b = 1 + p_\ell$, and the proof of
Proposition~\ref{cor:pLCS-k-emission-horizon} carries over, pLCS being
effective pWSTS~\cite{fffy2026}. This yields a computable depth that
grows only logarithmically in the total channel length. Since pLCS are
finitely branching, running $\mathcal{A}_2$, or any classical
path-enumeration algorithm~\cite{IyerNarasimha1997,AbdullaBenHendaMayr2007},
up to this depth bounds the number of explored states before the
exploration starts.

\begin{example}[Bursty router]
\label{ex:nu-pLCS-router}
A router receives traffic in bursts over an unreliable link. We model it
by a $\nu$-pLCS with $Q = \{\mathit{norm}, \mathit{fail}, \mathit{closed}\}$,
a single channel $\mathsf{buf}$,
$\Sigma = \{\mathit{data}, \mathit{bad}, \mathit{fin}\}$, where
$\mathit{bad}$ stands for a malformed packet,
$p_\ell \in (0,1) \cap \mathbb{Q}$, and $p(\nu)$ the Poisson distribution
of parameter $\nu \in \mathbb{Q}_{>0}$, which satisfies the assumptions of
Theorem~\ref{theoreme:nu-pLCS-effectiveness} and
Proposition~\ref{cor:pLCS-k-emission-horizon}. The rules, with arbitrary
positive weights, are a burst arrival
$\mathit{norm} \xrightarrow{\mathsf{buf}\,!\,\nu} \mathit{norm}$, the
processing of data
$\mathit{norm} \xrightarrow{\mathsf{buf}\,?\,\mathit{data}} \mathit{norm}$,
a failure $\mathit{norm} \xrightarrow{\mathsf{buf}\,?\,\mathit{bad}} \mathit{fail}$
when a malformed packet is processed, and a closing
$\mathit{norm} \xrightarrow{\mathsf{buf}\,?\,\mathit{fin}} \mathit{closed}$,
with self-loops $\mathit{fail} \xrightarrow{\tau} \mathit{fail}$ and
$\mathit{closed} \xrightarrow{\tau} \mathit{closed}$. A burst enters the
buffer in a single step: with a single-message emission model such as
pLCS, its first packets could be lost before the last ones arrive, which
would change the probabilistic behaviour. The failure states form the upward-closed set
$U = \{\mathit{fail}\} \times \Sigma^*$, and
$\widetilde{U} = \{\mathit{closed}\} \times \Sigma^*$. From
$x_0 = (\mathit{norm}, \epsilon)$, a burst starting with $\mathit{bad}$
can lead to $U$, whereas a burst starting with $\mathit{fin}$ can lead to
$\widetilde{U}$, so $0 < \mathbb{P}_{x_0}(\tau_U < \infty) < 1$. Every
configuration in $\mathit{norm}$ has infinitely many successors, so
classical path-enumeration
algorithms~\cite{IyerNarasimha1997,AbdullaBenHendaMayr2007} do not apply,
and neither does the GR-attractor
argument~\cite{AbdullaBenHendaMayrSandbergATVA2006}, since a step may add
several messages. Moreover, $U \cup \widetilde{U}$ is not uniformly
geometrically reachable: from $(\mathit{norm}, \mathit{data}^L)$, neither
$U$ nor $\widetilde{U}$ can be entered before the backlog has been
processed or lost, which, as in Example~\ref{ex:pLCS-not-coarse}, takes
arbitrarily long as $L$ grows. By
Proposition~\ref{cor:pLCS-k-emission-horizon}, an \textit{a priori}
computable depth in $O\big(\log(1/\varepsilon) + \log(|x| + 1)\big)$
from any configuration $x$, hence in $O(\log(1/\varepsilon))$ from $x_0$,
suffices for $\mathcal{A}_2$ to compute $\mathbb{P}_x(\tau_U < \infty)$
up to a precision $\varepsilon$
(see Corollary~\ref{prop:quant-from-horizon}).
\end{example}

\section{Conclusion}
\label{sec:conclusion}

Decisiveness guarantees that path exploration terminates, but not how
deep it must go. We introduced truncation sets, the horizons sufficient for
an $\varepsilon$-approximation, which are all non-empty, if and only if, the chain is decisive. In that case, our procedures compute an element of these
sets, and their least element in some cases, for instance for pVAS and
pLCS. These procedures, however, obtain a horizon only by exploring the
chain. Geometrically reachable sets provide such horizons \emph{a priori},
as functions of the precision and the initial state, whose evaluation,
for effective pWSTS, requires a constant number of elementary operations.
Two regimes appear. When the geometric bound is uniform in the initial
state, which is equivalent to bounded coarseness, horizons are
$O(\log(1/\varepsilon))$ for all initial states, as for pVAS and
Galton--Watson processes. Otherwise, a finite geometrically reachable
set, obtained from a drift condition, yields horizons in
$O(\log(1/\varepsilon) + \log w(x))$ from any initial state $x$. This
covers pLCS, which admit no horizon valid for all initial states, and
$\nu$-pLCS, whose infinite branching is out of reach of classical
path-enumeration algorithms~\cite{IyerNarasimha1997,AbdullaBenHendaMayr2007}.

Two directions remain open. The first is to identify a broader class of
effective Markov chains, strictly containing effective pWSTS, on which
these horizons remain computable. The second is to go beyond geometric
decay: chains whose tails only decrease polynomially, for which no
geometric drift is available, could still admit \emph{a priori} horizons,
now polynomial in $1/\varepsilon$.


\bibliographystyle{splncs04}
\bibliography{quantitative}

\newpage
\appendix

\section*{Use of AI.}
The authors take full responsibility for the entire content of this
paper, regardless of the assistance received from AI tools. The decision
to study decisiveness from a quantitative point of view, through
truncation sets and geometrically reachable sets, is our own, and was not
suggested by an AI. The same holds for the choice of the objects in their
present form, for the use of drift conditions to build finite
geometrically reachable sets, and for the decision to introduce
$\nu$-pLCS. The paper itself was written by the authors, albeit with
assistance.

We used large language models, namely Claude, Gemini, ChatGPT and
Perplexity, as follows. They helped explore various directions, some
fruitful and some not. The design of some examples, such as the bursty
router of Example~\ref{ex:nu-pLCS-router}, and of some proofs, among
them that of Theorem~\ref{thm:finite-attractor-pWSTS}, was AI-assisted,
through an iterative process in which the authors directed the
exploration, selected and adapted the suggestions, and designed and
validated the final versions. These tools also helped check our claims
against the literature on eagerness and decisiveness, and contributed to
the \LaTeX{} writing and condensing of the statements, proofs and proof
sketches. All proofs were reviewed by AI models, and all were checked by
the authors, AI review serving only as an additional verification.
Finally, the paper as a whole was proofread with AI assistance, the final
reading being carried out by the authors.

\section{Preliminaries}

\subsection{Markov chains and probabilistic WSTS}
\subsubsection{Selected pWSTS Models}

\label{annexe-pWSTS-def-models}
We recall the definitions of pVAS, pLCS, and multi-type Galton--Watson processes. These definitions are standard and follow those established in \cite{fffy2026}, allowing these models to be viewed as pWSTS.

\begin{definition}[VAS]\label{def:VAS}
    Let $d \in \Nat_+$, a $d$-VAS (or a VAS of dimension $d$) is a finite set $V\subseteq \mathbb {Z} ^{d}$.
\end{definition}

Let $d \in \Nat_+$ and let $V\subseteq \mathbb {Z} ^{d}$ be a d-VAS. For $u \in \mathbb N^d$ and $v\in V$, we write $u \xrightarrow{v} u+v$ whenever $u+v\in\mathbb N^d$. Moreover, let $w: V \to \mathbb{N}^+$ be a weight function. We define
transition probabilities as follows: for every configuration $u \in
\mathbb{N}^d$, let $\mathrm{En}(u) = \{v \in V: u + v \in \mathbb{N}^d\}$
denote its set of enabled actions. If $\mathrm{En}(u) \neq \emptyset$, then
for $v \in \mathrm{En}(u)$,
$$
  P(u)(u + v) = \frac{w(v)}{\sum_{v' \in \mathrm{En}(u)} w(v')},
$$
and $P(u)(x) = 0$ for every other state $x$. If $\mathrm{En}(u) =
\emptyset$, then $u$ is a deadlock and we set $P(u)(u) = 1$. Thus, $P$ defines a Markov kernel, and a d-VAS equipped with it is a pWSTS, since a d-VAS is a WSTS.

\begin{definition}[pLCS]
\label{def:pLCS}
A pLCS is a tuple $\mathcal{S} = (Q, \Sigma, \Delta, W, p_\ell)$ where:
\begin{itemize}
    \item $Q$ is a finite set of control states;
    \item $\Sigma$ is a finite alphabet of messages; 
    \item $\Delta \subseteq Q \times (\{!a, ?a \mid a \in \Sigma\} \cup \{\tau\}) \times Q$ is a finite set of transition rules; 
    \item $W: \Delta \to \mathbb{N}_+$ assigns a positive integer weight to each transition rule;
    \item $p_\ell \in (0,1)\cap \mathbb{Q}$ is a message loss probability.
\end{itemize}
\end{definition}

The configuration is a pair $(q,w)\in Q\times\Sigma^*$, where $w$ represents the contents of the FIFO channel. The state set is therefore $S = Q \times \Sigma^*$. We equip $\Sigma^*$ with the subword ordering relation $\sqsubseteq$, which is extended to configurations by $(q,w)\sqsubseteq(q',w') \iff q=q'$ and $w\sqsubseteq w'$. By Higman's Lemma, this ordering is a well-quasi-order on $S$. A transition is defined as, for $(q,w),(q',w') \in Q\times\Sigma^*$, $(q,w) \to (q',w')$ if and only if there exists $w'' \in \Sigma^*$ such that $(q,w) \to_{loss} (q,w'')$ and there exists $t \in (\{!a, ?a \mid a \in \Sigma\} \cup \{\tau\} )$ such that: $(q,t,q') \in \Delta$ and $(q,w'') \xrightarrow{t} (q',w')$. Where: \begin{itemize}
    \item $\to_{loss}$: for $(q,w),(q',w') \in Q\times\Sigma^*$, $(q,w) \to_{loss} (q',w')$, if and only if: $q = q'$ and $w' \sqsubseteq w$. This transition rule represents the loss of messages in the FIFO channel. Following \cite{ABDULLA2005141}, each message can be lost independently with probability $p_l$.
    \item for $q,q' \in Q$ and $a \in \Sigma$, $q \xrightarrow{!\,a} q'$ (\emph{emission}): always enabled, it appends the message $a$ to the tail of the channel, and moves to control state $q'$;
  \item for $q,q' \in Q$, $a \in \Sigma$ and $w \in \Sigma^*$, $q \xrightarrow{?\,a} q'$ (\emph{reception}): enabled iff $w$
  starts with $a$; it removes the head of the channel and moves to $q'$;
  \item for $q,q' \in Q$, $q \xrightarrow{\tau} q'$ (\emph{internal action}): always enabled;
  it moves to $q'$ without modifying the channels.
\end{itemize}

We assume that every control state has an outgoing emission or internal
rule. These rules being always enabled, pLCS are then deadlock-free. As shown in \cite{ABDULLA2005141}, pLCS can be extended with transition probabilities using the function $W$ and the parameter $p_\ell$. This probabilistic extension yields a Markov chain, which can formally be viewed as a pWSTS since the underlying lossy channel system with \emph{loss-action} semantics is already a WSTS.

\begin{definition}[Multi-type Galton--Watson process]
\label{def:Galton--Watson}
Let $d \in \Nat_+$ and for each $i \in \{1,...,d\}$, let $\mu^{(i)}$ be a probability distribution on $\Nat^d$. Let $(Z_n)_{n \in \Nat}$ in $\Nat^d$ be a stochastic process. $(Z_n)_{n \in \Nat}$
is a \emph{multi-type Galton--Watson process} associated with the family of laws of reproduction $\mu = (\mu^{(i)})_{1 \leq i \leq d}$ if:
$$
\begin{cases}
Z_0 = z_0 \in \Nat^d, \\[4pt]
Z_{n+1}^{(j)} = \displaystyle\sum_{i=1}^{d} \left( \sum_{k=1}^{Z_n^{(i)}} Y_{n,k}^{(i,j)} \right) \quad \text{for all } j \in \{1,..,d\} \text{ and } n \in \Nat,
\end{cases}
$$
where $(Y_{n,k}^{(i)})_{(n,k,i) \in \Nat\times\Nat_+ \times \{1,..,d\}}$ is a family of independent random vectors such that $Y_{n,k}^{(i)} = (Y_{n,k}^{(i,1)}, \ldots, Y_{n,k}^{(i,d)})$ is distributed according to $\mu^{(i)}$ for all $n\in \Nat$, $k \in \Nat_+$ and $i \in \{1,..,d\}$.

\end{definition}

The transition system is defined by $S = \Nat^d$ and, for $x, y \in \Nat^d$,
$x \to y$ iff $\mathbb{P}_x(Z_1 = y) > 0$. We equip $S$ with the
componentwise order. The system $(S, \leq, \to, P)$ associated with a multi-type Galton--Watson
process is a pWSTS.

\subsubsection{Procedures from \cite{fffy2026}}
\label{annexe-proc}

\label{annexe-proc-1}
\begin{algorithm}[H]
\caption{Coverability run construction for an effective pWSTS ($\mathcal{A}_{1}$)}
\label{alg:witness-path}
\begin{algorithmic}[1]
\Require $x \in S$, $\mathrm{In}_U$, $\mathrm{succ}$, $P$
\Ensure $\emptyset$ and $0$ if $x \notin \Pred^*(U)$; otherwise a finite run
$y_0 \cdots y_k$ with $y_0 = x$ and $y_k \in U$, along with its probability
$p^*$
\If{$\mathrm{In}_U(x,-1) = 0$} \Return $\emptyset, 0$ \EndIf
\State $k \gets 0$
\While{$\mathrm{In}_U(x, k) = 0$} \State $k \gets k + 1$ \EndWhile
\If{$k = 0$} \Return $x, 1$ \EndIf
\State $y_0 \gets x$; \; $p^* \gets 1$
\For{$i \gets 0$ \textbf{to} $k - 1$}
  \State $m \gets 1$; \; $z \gets \mathrm{succ}(y_i,m)$
  \While{$\mathrm{In}_U(z, k - i - 1) = 0$}
    \State $m \gets m + 1$; \; $z \gets \mathrm{succ}(y_i,m)$
  \EndWhile
  \State $y_{i+1} \gets z$; \; $p^* \gets p^* \times P(y_i)(y_{i+1})$
\EndFor
\State \Return $y_0\, y_1 \cdots y_k,\ p^*$
\end{algorithmic}
\end{algorithm}

\label{annexe-proc-2}
\begin{algorithm}[H]
\caption{Bounded-horizon approximation ($\mathcal{A}_{2}$)}
\label{alg:bounded-horizon}
\begin{algorithmic}
\Require $B$, $B'$, $\mathrm{In}$, $x \in S$, $\ell \in \Nat$,
$\varepsilon \in \mathbb{Q} \cap (0,1)$, $\mathrm{succ}$, $P$
\Ensure $p^\ast \in \mathbb{Q}$ such that
$p^\ast \leq \mathbb{P}_x(\tau_{ {\uparrow}B} \leq \ell) \leq p^\ast + \varepsilon$
\If{$\ell = 0$} \Return $\mathrm{In}(x, B)$ \EndIf
\State $p^\ast \gets 0$; \; $\gamma \gets \varepsilon / \ell$
\State $F \gets \{\, x \mapsto 1 \,\}$
  \Comment{frontier at depth $0$: a finite map from states to probabilities}
\For{$k = 0$ \textbf{to} $\ell - 1$}
  \State $F' \gets \emptyset$
  \ForAll{$(y, p) \in F$}
    \If{$\mathrm{In}(y, B) = 1$} \State $p^\ast \gets p^\ast + p$
    \ElsIf{$\mathrm{In}(y, B') = 1$}
      \State $q \gets 0$; \; $m \gets 1$
      \State \textbf{while} $q < 1 - \gamma$ \textbf{do}
      \State \quad $z \gets \mathrm{succ}(y, m)$; \;
      $q \gets q + P(y)(z)$; \; $m \gets m + 1$
      \State \quad $F'[z] \gets F'[z] + p \cdot P(y)(z)$
        \Comment{merge; absent keys read as $0$}
      \State \textbf{end while}
    \EndIf
  \EndFor
  \State $F \gets F'$
\EndFor
\ForAll{$(y, p) \in F$}
  \Comment{last level: no further expansion}
  \If{$\mathrm{In}(y, B) = 1$} \State $p^\ast \gets p^\ast + p$ \EndIf
\EndFor
\State \Return $p^\ast$
\end{algorithmic}
\end{algorithm}

\label{annexe-proc-3}
\begin{algorithm}[H]
\caption{Infinite horizon approximation ($\mathcal{SA}_{3}$)}
\label{alg:dynamic-truncation}
\begin{algorithmic}[1]
\Require $B$, $B'$, $\mathrm{In}$, $x \in S$,
$\varepsilon \in \mathbb{Q} \cap (0,1)$, $\mathrm{succ}$, $P$  
\Ensure $p^\ast \in \mathbb{Q}$ such that
$p^\ast \leq \mathbb{P}_x(\tau_{\uparrow B} < \infty) \leq p^\ast + \varepsilon$
\State $p^\ast \gets 0$; \; $k \gets 0$
\State $F \gets \{\, x \mapsto 1 \,\}$
  \Comment{frontier at depth $k$: a finite map from states to probabilities}
\While{$\sum_{(y,p) \in F} p > \varepsilon/2$}
  \State $\gamma_k \gets \varepsilon / 2^{\,k+2}$
    \Comment{$\sum_{k\ge0} \varepsilon\,2^{-(k+2)} = \varepsilon/2$:
    accumulated branching error bounded by $\varepsilon/2$}
  \State $F' \gets \emptyset$
  \ForAll{$(y, p) \in F$}
    \If{$\mathrm{In}(y, B) = 1$} \State $p^\ast \gets p^\ast + p$
    \ElsIf{$\mathrm{In}(y, B') = 1$}
      \State $q \gets 0$; \; $m \gets 1$
      \State \textbf{while} $q < 1 - \gamma_k$ \textbf{do}
      \State \quad $z \gets \mathrm{succ}(y, m)$; \;
      $q \gets q + P(y)(z)$; \; $m \gets m + 1$
      \State \quad $F'[z] \gets F'[z] + p \cdot P(y)(z)$
        \Comment{merge; absent keys read as $0$}
      \State \textbf{end while}
    \EndIf
  \EndFor
  \State $k \gets k + 1$
  \State $F \gets F'$
\EndWhile
\State \Return $p^\ast$
\end{algorithmic}
\end{algorithm}

\section{Geometric reachability}

\subsection{Computing an element of $n^*_{\uparrow B}$}

\paragraph{Remark~\ref{rem:proc3}}

\label{annexe-proof-n-star-remark}
\begin{lemma}
\label{lem:proc3-depth}
Assume that $(\mathcal{S}, P)_{\mathrm{eff}}$ is decisive with respect to ${\uparrow}B$, and let $K$ be the depth at which $\mathcal{SA}_3$ exits its main loop. Then $\max(0,K - 1) \in n^*_{{\uparrow}B}(\varepsilon, x)$.
\end{lemma}

\begin{proof}
Let $U = {\uparrow}B$. For $k \le K$, let $F_k$ be the frontier at depth $k$, $m_k = \sum_{(y,p) \in F_k} p$ its mass, and $T_k = \mathbb{P}_x(\tau_{U \cup \widetilde{U}} \geq k)$. Note that $K \geq 1$, since $m_0 = 1 > \varepsilon/2$. As in~\cite{fffy2026}, we write $a^{(y)}_z$ for the rational lower bound of $P(y)(z)$ actually used by $\mathcal{SA}_3$.

We first show by induction on $k \le K$ that for every $z \in S$:
$
  F_k[z] \le \mathbb{P}_x(X_k = z,\ \tau_{U \cup \widetilde{U}} \geq k) 
$
where absent keys of $F_k$ are read as $0$. For $k = 0$, we have $F_0 = \{x \mapsto 1\}$ and $\mathbb{P}_x(X_0 = x,\ \tau_{U \cup \widetilde{U}} \geq 0) = 1$, while both sides vanish for $z \neq x$.

For $k \in \Nat$, assume the inequality holds at depth $k$. By merging, each state appears at most once in $F_k$. Moreover, during iteration $k$, an entry $(y, p) \in F_k$ is expanded if and only if $y \notin U \cup \widetilde{U}$, and in that case each enumerated successor $z$ receives the contribution $p\, a^{(y)}_z$, with $a^{(y)}_z \le P(y)(z)$. Successors that are not enumerated receive nothing. Hence, for every $z \in S$,
$$
  F_{k+1}[z]
  \le \sum_{\substack{(y,p) \in F_k \\ y \notin U \cup \widetilde{U}}} p\, P(y)(z)
  \le \sum_{y \notin U \cup \widetilde{U}} \mathbb{P}_x(X_k = y,\ \tau_{U \cup \widetilde{U}} \geq k)\, P(y)(z),
$$
where the second inequality uses the induction hypothesis and the non-negativity of all terms. The event $\{X_k = y,\ \tau_{U \cup \widetilde{U}} \geq k\}$ belongs to $\mathcal{F}_k$, so the Markov property (Proposition~\ref{prop:markov}), applied with $\Phi = \mathbf{1}_{\{X_1 = z\}}$, gives
$
  \mathbb{P}_x(X_k = y,\ \tau_{U \cup \widetilde{U}} \geq k,\ X_{k+1} = z) = \mathbb{P}_x(X_k = y,\ \tau_{U \cup \widetilde{U}} \geq k)\, P(y)(z).
$
Since $\{\tau_{U \cup \widetilde{U}} \geq k+1\} = \{\tau_{U \cup \widetilde{U}} \geq k\} \cap \{X_k \notin U \cup \widetilde{U}\}$, summing over $y \notin U \cup \widetilde{U}$ yields
$
  F_{k+1}[z] \le \mathbb{P}_x(X_{k+1} = z,\ \tau_{U \cup \widetilde{U}} \geq k+1),
$
which completes the induction.

Summing over $z \in S$, we obtain $m_k \le T_k$, so $D_k = T_k - m_k \geq 0$, with $D_0 = 0$. We now bound the growth of $D_k$. On the one hand, using the same decomposition of $\{\tau_{U \cup \widetilde{U}} \geq k+1\}$ and the non-negativity of each difference $\mathbb{P}_x(X_k = y,\ \tau_{U \cup \widetilde{U}} \geq k) - F_k[y]$,
$$
  T_{k+1} = \sum_{y \notin U \cup \widetilde{U}} \mathbb{P}_x(X_k = y,\ \tau_{U \cup \widetilde{U}} \geq k)
  \le \sum_{\substack{(y,p) \in F_k \\ y \notin U \cup \widetilde{U}}} p \; + \; D_k .
$$
On the other hand, the inner loop at state $y$ exits with an accumulated mass $q_y \geq 1 - \gamma_k$, and $m_k \le 1$, so that
$
  m_{k+1} = \sum_{\substack{(y,p) \in F_k \\ y \notin U \cup \widetilde{U}}} p\, q_y
  \geq \sum_{\substack{(y,p) \in F_k \\ y \notin U \cup \widetilde{U}}} p \; - \; \gamma_k .
$
Subtracting the two bounds gives $D_{k+1} \le D_k + \gamma_k$, hence $D_K \le \sum_{k<K} \gamma_k \le \varepsilon/2$. The exit condition of the main loop gives $m_K \le \varepsilon/2$. Therefore
$
  \mathbb{P}_x(\tau_{U \cup \widetilde{U}} > K-1) = T_K = m_K + D_K \le \varepsilon,
$
that is, $K - 1 \in n^*_{{\uparrow}B}(\varepsilon, x)$.
\end{proof}

We denote by $(\mathcal{SA}_3)'$ the procedure that runs $\mathcal{SA}_3$ and returns $K - 1$, where $K$ is the depth at which $\mathcal{SA}_3$ exits its main loop. 

\begin{proposition}
    If $(\mathcal{S}, P)_{\mathrm{eff}}$ is decisive with respect to ${\uparrow}B$, $(\mathcal{SA}_3)'$ returns an element of $n^*_{{\uparrow}B}(\varepsilon, x)$.
\end{proposition}

\begin{proof}
    If $(\mathcal{S}, P)_{\mathrm{eff}}$ is decisive with respect to ${\uparrow}B$, then $\mathcal{SA}_3$ terminates~\cite{fffy2026}, and Lemma~\ref{lem:proc3-depth} ensures that $(\mathcal{SA}_3)'$ returns an element of $n^*_{{\uparrow}B}(\varepsilon, x)$.
\end{proof}

\subsection{Geometrically reachable sets}
\label{app:eagerness}

The first definition of eagerness~\cite{AbdullaBenHendaMayrSandbergATVA2006} reads as follows: $(S,P)$ is \emph{eager with respect to $x \in S$} for a target $F$ if there exist $\beta_x < 1$ and $N_x \in \Nat$ such that $\mathbb{P}_x(n \le \tau_F < \infty) \le \beta_x^n$ for all $n \geq N_x$. It is equivalent to $\{x\}$-eagerness. Indeed, if $\mathbb{P}_x(n \le \tau_F < \infty) \leq C_x (\beta'_x)^n$ for all $n \geq 0$, then for any $\beta_x \in (\beta'_x, 1)$ the bound $\beta_x^n$ holds for all $n \geq N_x = \max\big(0, \lceil \log C_x / \log(\beta_x/\beta'_x) \rceil\big)$. Conversely, if the bound $\beta_x^n$ holds for $n \geq N_x$ with $\beta_x > 0$, then $C_x = \beta_x^{-N_x}$ gives the bound $C_x \beta_x^n$ for all $n \geq 0$.

Earlier versions of eager attractors~\cite{AbdullaBenHendaMayrSandbergATVA2006,AbdullaBenHendaMayrSandbergQEST2006} state the bound as $\beta^n$, without the factor $C$. Since the bound is required for all $n \geq 0$, this is not a mere reformulation: for $n = 1$, it forces $\mathbb{P}_x(X_1 \notin A) \le \beta < 1$ for every $x \in A$, which fails for instance for $A = \{a\}$ in the finite chain $a \to b \to a$.

\end{document}